\documentclass[twocolumn]{aastex63}
\usepackage{amsmath}
\usepackage{comment}
\usepackage{graphicx}
\usepackage{xcolor}
\usepackage{xparse,xcoffins}

\received{May 1, 2021}
\revised{July 1, 2021}
\accepted{\today}

\submitjournal{APJ}

\shorttitle{$L(M)$ model}
\shortauthors{Yang et al.}

\usepackage{tgtermes}

\newcommand{\ci}{\ion{C}{1}}
\newcommand{\cii}{\ion{C}{2}}

\newcommand{\oiii}{\ion{O}{3}}
\newcommand{\nii}{\ion{N}{2}}

\ExplSyntaxOn
\NewCoffin\imagecoffin
\NewCoffin\labelcoffin

\keys_define:nn { miguel/label }
 {
  label   .tl_set:N = \l_miguel_label_tl,
  labelbox .bool_set:N = \l_miguel_label_box_bool,
  labelbox .default:n = true,
  fontsize .tl_set:N = \l_miguel_label_size_tl,
  fontsize .initial:n = \footnotesize,
  pos .choice:,
  pos/nw .code:n = \tl_set:Nn \l_miguel_label_pos_tl { left,up },
  pos/ne .code:n = \tl_set:Nn \l_miguel_label_pos_tl { right,up },
  pos/sw .code:n = \tl_set:Nn \l_miguel_label_pos_tl { left,down },
  pos/se .code:n = \tl_set:Nn \l_miguel_label_pos_tl { right,down },
  pos/n .code:n = \tl_set:Nn \l_miguel_label_pos_tl { hc,up },
  pos/w .code:n = \tl_set:Nn \l_miguel_label_pos_tl { left,vc },
  pos/s .code:n = \tl_set:Nn \l_miguel_label_pos_tl { hc,down },
  pos/e .code:n = \tl_set:Nn \l_miguel_label_pos_tl { right,vc },
  pos .initial:n = nw,
  unknown .code:n   = \clist_put_right:Nx \l_miguel_label_clist
                       { \l_keys_key_tl = \exp_not:n { #1 } }
 }
\clist_new:N \l_miguel_label_clist
\box_new:N \l_miguel_label_box
\box_new:N \l_miguel_label_image_box

\NewDocumentCommand{\xincludegraphics}{O{}m}
 {
  \group_begin:
  \tl_clear:N \l_miguel_label_tl
  \clist_clear:N \l_miguel_label_clist
  \keys_set:nn { miguel/label } { #1 }
  \tl_if_empty:NTF \l_miguel_label_tl
   {
    \miguel_includegraphics:Vn \l_miguel_label_clist { #2 }
   }
   {
    \SetHorizontalCoffin\imagecoffin
     {
      \miguel_includegraphics:Vn \l_miguel_label_clist { #2 }
     }
    \SetHorizontalCoffin\labelcoffin
     {
      \raisebox{\depth}
       {
        \bool_if:NTF \l_miguel_label_box_bool
         { \fcolorbox{white}{white}{\l_miguel_label_size_tl\l_miguel_label_tl} }
         { \l_miguel_label_size_tl\l_miguel_label_tl }
       }
     }
    \SetVerticalPole\imagecoffin{left}{3pt+\CoffinWidth\labelcoffin/2}
    \SetVerticalPole\imagecoffin{right}{\Width-3pt-\CoffinWidth\labelcoffin/2}
    \SetHorizontalPole\imagecoffin{up}{\Height-3pt-\CoffinHeight\labelcoffin/2}
    \SetHorizontalPole\imagecoffin{down}{3pt+\CoffinHeight\labelcoffin/2}
    \use:x{\JoinCoffins\imagecoffin[\l_miguel_label_pos_tl]\labelcoffin[vc,hc]} 
    \TypesetCoffin\imagecoffin
   }
   \group_end:
 }
\NewDocumentCommand{\setlabel}{m}
 {
  \keys_set:nn { miguel/label } { #1 }
 }

\cs_new_protected:Nn \miguel_includegraphics:nn
 {
  \includegraphics[#1]{#2}
 }
\cs_generate_variant:Nn \miguel_includegraphics:nn { V }

\ExplSyntaxOff

\begin{document}


\title{An empirical representation of a physical model for the ISM [\cii], CO, and [\ci] emission at redshift $1\leq z\leq9$}

\correspondingauthor{Shengqi Yang}
\email{syang@carnegiescience.edu}

\author{Shengqi Yang}
\affiliation{Carnegie Observatories, 813 Santa Barbara Street, Pasadena, CA 91101, U.S.A}
\affiliation{Center for Cosmology and Particle Physics, Department of Physics, New York University, 726 Broadway, New York, NY, 10003, U.S.A.}

\author{Gerg\"o Popping}
\affiliation{European Southern Observatory, Karl-Schwarzschild-Strasse 2, D-85748, Garching, Germany}

\author{Rachel S. Somerville}
\affiliation{Center for Computational Astrophysics, Flatiron Institute, New York, NY 10010, U.S.A.}

\author{Anthony R. Pullen}
\affiliation{Center for Cosmology and Particle Physics, Department of Physics, New York University, 726 Broadway, New York, NY, 10003, U.S.A.}
\affiliation{Center for Computational Astrophysics, Flatiron Institute, New York, NY 10010, U.S.A.}

\author{Patrick C. Breysse}
\affiliation{Center for Cosmology and Particle Physics, Department of Physics, New York University, 726 Broadway, New York, NY, 10003, U.S.A.}
\affiliation{Center for Computational Astrophysics, Flatiron Institute, New York, NY 10010, U.S.A.}

\author{Abhishek S. Maniyar}
\affiliation{Center for Cosmology and Particle Physics, Department of Physics, New York University, 726 Broadway, New York, NY, 10003, U.S.A.}
 
\begin{abstract}
Sub-millimeter emission lines produced by the interstellar medium (ISM) are strong tracers of star formation and are some of the main targets of line intensity mapping (LIM) surveys. In this work we present an empirical multi-line emission model that simultaneously covers the mean, scatter, and correlations of [\cii], CO J=1-0 to J=5-4, and [\ci] lines in redshift range $1\leq z\leq9$. We assume the galaxy ISM line emission luminosity versus halo mass relations can be described by double power laws with redshift-dependent log normal scatter. The model parameters are then derived by fitting to the state of the art semi-analytic simulation results that have successfully reproduced multiple sub-millimeter line observations at $0\leq z\lesssim6$. We cross check the line emission statistics predicted by the semi-analytic simulation and our empirical model, finding that at $z\geq1$ our model reproduces the simulated line intensities with fractional error less than about 10\%. The fractional difference is less than 25\% for the power spectra. Grounded on physically-motivated and self-consistent galaxy simulations, this computationally efficient model will be helpful in forecasting ISM emission line statistics for upcoming LIM surveys.

\end{abstract}


\section{Introduction} \label{sec:intro}
Two cosmic epochs that are of great interest for modern astrophysics are the epoch of reionization (EoR) at redshift $z>6$ and the cosmic ``high noon" era, when the cosmic star formation history peaked, at $z\sim2$ \citep{2014ARA&A..52..415M}. Since the molecular clouds distributed throughout the interstellar medium (ISM) host star formation, and the star-forming activity in return influences the nearby ISM environment, ISM emission lines are unique tracers for the physical properties of the stellar population as well as the star-forming environment. Among various molecular and fine structure lines, radiation with wavelength between 15 $\mu$m and 1 mm, also referred to as sub-millimeter emission lines, are particularly promising for observation since they can penetrate dense molecular clouds. Some of the popular sub-millimeter observational targets are the molecular ISM tracers, including the CO rotational transitions, and [\ci] fine structure lines, tracers of photodissociation regions such as the [\cii] 157 $\mu$m line, which is also the brightest far-infrared line, and tracers of HII regions, including the [\oiii] 88 $\mu$m and 52 $\mu$m fine structure lines, [\nii] 122 $\mu$m and 205 $\mu$m transitions.\par
Our observational view of sub-mm emission line properties of galaxies in the nearby and distant Universe has exploded since the launch of the Herschel Space telescope and the commissioning of the Atacama Large Millimeter Array (ALMA) in 2010. However, there is concern about whether the bright sources observable by those instruments are representative of the general galaxy population. An emerging observational technique referred to as line intensity mapping (LIM) \citep{2017arXiv170909066K} is designed to complement traditional galaxy surveys. Instead of resolving individual targets with high spatial resolution, LIM measures the aggregated emission along the line of sight with high spectral resolution, including radiation contributed by faint galaxies. LIM is therefore more suitable for probing the average galaxy properties during the cosmic epochs of interest. However, since the beam sizes of LIM surveys are generally much larger than the target sizes, confusion is a major challenge for LIM survey data analysis \citep{1990LIACo..29..117H}.\par
Up to now, LIM surveys have achieved preliminary [\cii] and CO detections at high redshift \citep{2018MNRAS.478.1911P,2019MNRAS.489L..53Y,Keating2016,2020ApJ...901..141K}, and more surveys with higher spectral resolution are planned in the upcoming decade to further confirm or extend those detections. Motivated by the prospects of these upcoming experiments, reliable sub-millimeter line emission models are needed to provide theoretical predictions for the observational interpretation as well as forecasts. Ideally, such a sub-mm model framework should connect the line emission observables to the physical properties of the ISM gas clouds, galaxies, and dark matter halos. It should also simultaneously predict as many emission lines as possible to provide a comprehensive understanding of the multi-phase ISM environment. It is difficult to construct such an analytic model because the micro-physical processes involved in different sub-millimeter line emissions vary case by case. Moreover, it is challenging to connect line emission mechanisms happening on the atomic scale to the sub-pc scale molecular clouds, kpc scale galaxies, and Mpc scale dark matter halos. There are many analytic models in the literature describing the major sub-millimeter LIM targets, including CO, [\cii], [\oiii], and [\nii] lines (e.g. \cite{2011ApJ...741...70L,2012ApJ...745...49G,2013ApJ...768...15P,2014ApJ...795..174K,2015ApJ...806..209S,2016ApJ...817..169L,2015ApJ...802...81M,2018MNRAS.478.1911P,2019MNRAS.488.3014P,2019ApJ...887..142S,2020MNRAS.499.3417Y,Padmanabhan2021}). All those models include empirical treatments to some degree, and fail to simultaneously cover all the lines of interest. In particular, there is a lack of models that predict the CO rotational transitions with high J quantum states, which are the major low-redshift interlopers for many LIM surveys.\par 
An alternative approach is to combine numerical hydrodynamic galaxy formation simulations with radiative transfer, photo-ionization, and photo-dissociation region modeling (e.g. \cite{Olsen:2017,2018MNRAS.481L..84M,2020ApJ...905..102L,Olsen:2021}). The strengths of the numerical simulation approach is that detailed properties of the gas distribution, temperature, and radiation field can be obtained down to the resolution of the simulation. However, all simulations that cover cosmological volumes have resolution considerably coarser than the molecular cloud scales relevant for line emission, so a ``sub-grid" model must be applied on top of the simulation as part of the line emission model. Due to the rather high computational expense of this approach, it is currently not feasible to produce predictions for large numbers of galaxies (state of the art is $< 700$ galaxies) or to explore uncertainties in the sub-grid recipes for physical processes in the simulations or in the line emission modeling interface. Also, portability of the numerical approach can be limited since many simulations are not open source. 

A ``middle way" is provided by coupling semi-analytic models of galaxy formation with a sub-mm line emission modeling framework \citep{Popping:2014b,Popping:2016,2019MNRAS.482.4906P,2018A&A...609A.130L}. 
The state of the art sub-mm line simulation framework is introduced in \cite{2019MNRAS.482.4906P} (hereafter Popping2019) and \cite{2021ApJ...911..132Y}. This framework combines the semi-analytic galaxy formation model (SAM) developed by the ``Santa Cruz" group \citep{1999MNRAS.310.1087S,2008MNRAS.391..481S,2012MNRAS.423.1992S,2014MNRAS.444..942P,2014MNRAS.442.2398P,2015MNRAS.453.4337S} with a sub-grid model for molecular cloud properties and the \textsc{DESPOTIC} \citep{2014MNRAS.437.1662K} spectral synthesis tool. The SAM provides information on the global properties (gas mass, metallicity, density) for up to millions of galaxies over a very wide range in halo mass and redshift, and pre-computed tables from \textsc{DESPOTIC} provide the calculations of the thermo-chemical evolution and emergent line emission from optically thick clouds. The models include physics-grounded treatments for ISM multi-phase partitioning, molecular cloud distribution, and cloud density profile variation. The full pipeline is many orders of magnitude faster than a fully numerical approach, and yields results that are broadly consistent \citep[e.g.][]{2020ApJ...905..102L}. 
The Popping2019 sub-mm-SAM framework has successfully reproduced the observed strength of [\cii] 157 $\mu$m fine structure lines, multiple CO rotational transitions, and [\ci] fine structure emissions at $0\leq z\lesssim6$. \par
Motivated by the need for a physically grounded multi-line analytic model, in this work we present an empirical model for [\cii], CO, and [\ci] line luminosity as a function of halo mass, calibrated to accurately reproduce the updated Popping2019 sub-mm-SAM results. We assume that the line intensity $L$ versus halo mass $M$ relations follow a double power law trend, together with a redshift-dependent log normal scatter. We then study the evolution of $L(M)$ relations as well as their scatter and correlations over the redshift range $0<z\leq9$. As a first application, we use this model to predict the line intensities and power spectra, finding that our empirical model yields results that match the direct SAM simulation results with fractional error better than 25\% at $1\leq z\leq9$. Given that CO rotational transitions and [\ci] fine structure lines are potentially important low redshift interlopers for many [\cii] LIM surveys, this model will be useful for testing the signal detectability and interloper line removal techniques for upcoming LIM surveys (e.g. \cite{2021arXiv210614904B,Pullen2022EXCLAIM}).\par
We emphasize that the Santa Cruz SAM contains numerous simplifying assumptions, and there are many uncertainties in the physical processes, particularly those connected to stellar and AGN feedback. This is the case for all existing \emph{a priori} models of galaxy formation, whether semi-analytic or numerical \citep[e.g.][]{SD15}. The model assumptions are based on phenomenology or results from numerical simulations, and the fiducial model parameters have been chosen to produce predictions that are consistent with a broad range of observations of galaxy properties, as documented in an extensive literature stretching over several decades (see Section~\ref{sec:SAM}). Similarly, there are also uncertainties and assumptions that affect the predictions of the Popping2019 sub-mm SAM model (which have been rather extensively documented in that work). We do not mean to suggest that these models are the definitive ``right answer'' -- instead, our goal is to introduce a framework for efficiently translating the predictions from physically motivated models into LIM observables. This framework is quite general, and should be extensible to nearly any physics-based model or to variations in the model physics modules and parameters. 

The plan of this paper is as follows: We introduce the Santa Cruz SAM for galaxy formation in section \ref{sec:SAM}. In section \ref{sec:submmSAM} we review the \textsc{DESPOTIC} based sub-mm model proposed by Popping2019 for estimating the sub-millimeter line luminosities of the simulated galaxies, and we describe the model refinement made in this work. We introduce the empirical $L(M)$ relations and the model calibration processes in section \ref{sec:model}. The SAM-model line statistics comparisons are presented in section \ref{sec:crosscheck}. We conclude in section \ref{sec:conclusion}. Throughout this work we adopt a flat $\Lambda$CDM cosmology and assume cosmological parameters $\Omega_m=0.308$, $\Omega_\Lambda=0.692$, $h=H_0/(100\mathrm{km/s/Mpc})=0.678$, $\sigma_8=0.831$, and $n_s=0.9665$ \citep{2016A&A...594A..13P}. We adopt a baryonic fraction of $f_b=0.1578$ and a Chabrier initial mass function \citep{2003PASP..115..763C}.\par   

\section{Santa Cruz semi-analytic model} \label{sec:SAM}
In this work we use the SAM developed by the ``Santa Cruz" group to generate galaxy samples. The development of this SAM framework is introduced in detail in a series of papers \citep{1999MNRAS.310.1087S,2008MNRAS.391..481S,2012MNRAS.423.1992S,2014MNRAS.444..942P,2014MNRAS.442.2398P,2015MNRAS.453.4337S}. We will only introduce the key features in this section and we refer our readers to \cite{2008MNRAS.391..481S, 2015MNRAS.453.4337S} for more details.\par
Instead of solving the equations of gravity, hydro-, and thermodynamics for particles or grid cells, SAMs adopt physically motivated treatments of the bulk flows of mass and metals between different ``reservoirs" (e.g., the diffuse intergalactic medium outside of resolved halos, hot diffuse halo gas, cold dense ISM gas, stars), set within the framework of dark matter halo formation predicted by $\Lambda$CDM and encapsulated in the form of ``merger trees". 

First, we divide the dark matter halo mass range that we consider in this work ($10\leq\log(M_\mathrm{halo}/[M_\odot])\leq13$)\footnote{In this work we use $\log$ to denote a base-10 logarithm, while $\ln$ denotes a natural logarithm.} into a hundred log mass bins and generate a hundred dark matter halos in each bin. This halo mass range is selected because halos with mass greater than $10^{13} M_\odot$ are very rare objects, while the SAM predicts that halos less massive than $10^{10} M_\odot$ are too faint to significantly influence the sub-mm line emission statistics (see \cite{2021ApJ...911..132Y} Appendix A for a more detailed discussion). The SAM then estimates the merging history of each halo using a method based on the extended Press-Schechter formalism \citep{1999MNRAS.305....1S,2008MNRAS.391..481S}. The halo merger tree is evolved back in time until the progenitors are less massive than one percent of the root halo or $10^{8}M_\odot$, whichever is smaller.  Before cosmic reionization, each halo is assigned hot gas with mass determined by the product of the Universal baryonic fraction and the halo virial mass. After reionization, the mass filtering scale associated with UV photoionization heating by the meta-galactic background from hydrodynamic simulations \citep{2008MNRAS.390..920O} is used to determine the fraction of gas that is accreted as a function of halo mass and redshift.

The SAM computes the rate that gas cools and accretes into the ISM using a spherically symmetric standard cooling model as described in \citet{2008MNRAS.391..481S}. Gas is assumed to initially form a disc with an exponential radial density profile, and the disc size is estimated following \citet{2008ApJ...672..776S}. The cold gas surface density profile is used along with the gas metallicity and the local UV radiation field (assumed proportional to the SFR) to estimate the fraction of ionized, atomic, and molecular gas in each annulus as a function of radius within each disc, using a prescription based on numerical hydrodynamic simulations \citep{2015MNRAS.453.4337S}. The new SFR is then computed based on the surface density of molecular gas, motivated by observations. This feature makes Santa Cruz SAM particularly suitable for ISM line emission predictions. The SFR, stellar mass, cold gas mass, and other predictions of the Santa Cruz SAM have been shown to be broadly consistent with observations from the local universe to $z\sim10$ \citep{2014MNRAS.442.2398P,2015MNRAS.453.4337S,2019MNRAS.483.2983Y,2019MNRAS.490.2855Y,2020MNRAS.496.4574Y,2020MNRAS.494.1002Y,somerville:2021}.\par

\section{Sub-mm emission line modeling}\label{sec:submmSAM}
In this work we adopt the sub-mm line-emission modeling framework proposed by Popping2019 (hereafter sub-mm SAM), which adds a sub-grid model for molecular cloud properties to the Santa Cruz SAM and couples this cloud population to the \textsc{DESPOTIC} code. As mentioned in the introduction, one major challenge of applying sub-mm line emission models to cosmological simulations is building connections between physical processes happening at atomic scales, molecular cloud scales, and the scales of galaxies. Popping2019 developed a sub-grid approach to address this problem. Specifically, each galaxy simulated by the SAM is divided into multiple radial annuli. In each ISM annulus, the masses of the molecular cloud population are assigned by selecting randomly from a power law mass distribution with parameters that depend on the mean density in the annulus. The molecular clouds are all assumed to be spherically symmetric and are assigned a radial density profile. The [\cii], CO, and [\ci] luminosities of each molecular cloud are then estimated through the following process: First, multiple grid points are selected in the five dimensional parameter space including cloud mass, gas pressure, metallicity, external ultra violet and cosmic ray radiation field strength,  and the redshift \{$M_\mathrm{cloud}$, $P_\mathrm{ext}$, $Z$, $F_\mathrm{ext}$, $z$\}. To ensure convergence of the numerical solutions, each molecular cloud is further divided into 25 regions so that the gas property within each region is close to constant. This set of physical properties is then input to \textsc{DESPOTIC}, which uses a one-zone model to compute the dominant heating, cooling, and chemical processes. These include grain photoelectric heating, heating of the dust by infrared and ultraviolet radiation, thermal cooling of the dust,  cosmic ray and X-ray heating, collisional energy exchange between dust and gas, and a simple network for carbon chemistry. The numerical solutions provided by \textsc{DESPOTIC} are used to create a lookup table, which allows a mapping from a set of physical properties to [\cii], CO, and [\ci] line luminosity. Finally, the relevant sub-mm line luminosities of each simulated galaxy is estimated by summing up the radiation from all the molecular clouds within it.\par

The sub-mm SAM line luminosity simulation results are somewhat sensitive to the assumed cloud radial density profile. Popping2019 have shown that models with a Plummer density profile reproduce various [\cii], CO, and [\ci] line observations at redshift $0\leq z\lesssim6$. However, several subsequent works have shown that the sub-mm SAM underestimates [\cii] and [\ci] luminosities compared to more recent ALMA detections at high redshift (e.g. \cite{2020ApJ...890...24V,2020A&A...643A...2B}). To resolve the tension between the sub-mm SAM predictions and recent observations, in this work we assume a power law radial density profile for each molecular cloud, as first suggested by \cite{2021ApJ...911..132Y}. This change of the cloud density profile does not significantly influence the CO luminosity predictions, but it effectively increases the [\cii] and [\ci] line strengths at high redshift such that the model predictions are consistent with observations. We also created a finer grid in parameter space for the \textsc{DESPOTIC} lookup table calculation to ensure accurate linear interpolations. We randomly select 115000 test points in the 5D parameter space \{$10^4<M_\mathrm{cloud}/[M_\odot]<10^7$, $10^3< P_\mathrm{ext}/k_\mathrm{B}/10^4[\mathrm{cm^{-3}K]}<10^9$, $10^{-3}< Z/[Z_\odot]<10^{0.5}$, $10^{-3}< F_\mathrm{ext}/[\mathrm{Habing}]<10^{4}$, $0<z<9$\} and find that more than 78.5\% of the [\cii], CO and [\ci] linear interpolations given by the lookup table agree with the \textsc{DESPOTIC} numerical solutions with better than 20\% fractional error. \par

\section{An empirical [\cii], CO, and [\ci] model calibrated to the sub-mm SAM}\label{sec:model}
In order to represent the sub-mm SAM simulation results in a more computationally efficient and portable way, in this section we introduce an empirical model that captures the average, dispersion, and correlation of line luminosities versus halo mass relations $L(M)$ simulated following the framework introduced in section \ref{sec:SAM} and section \ref{sec:submmSAM}. This model reproduces SAM line intensities and power spectra with better than 25\% fractional error at $z\geq1$. The SAM-model power spectra fractional difference further reduces to less than 15\% at $z\geq 3$. For applications that require more accurate sub-mm line simulations, in particular at redshift $z<1$, we provide tables of the sub-mm SAM $L(M)$ simulation results at \url{ https://users.flatironinstitute.org/~rsomerville/Data\_Release/LIM/}.\par
In this work we define the sub-mm line luminosity of each dark matter halo as the sum of the luminosity of all galaxies within the halo, and treat halos as the smallest sub-mm line emitters. The $L(M)$ model therefore cannot capture the sub-mm line fluctuations on scales smaller than the size of the halo. Since the beam sizes of LIM surveys are generally large, such a model is still suitable for making many LIM forecasts. We leave a more careful treatment of the ``one halo term" (which would require separate modeling of central and satellite galaxies, along with their location within the halo), which is crucial for galaxy spectroscopic survey forecasts, to future works.\par
\subsection{The average $L(M)$ model}\label{sec:LM}
\begin{figure}
    \centering
    \includegraphics[width=0.49\textwidth]{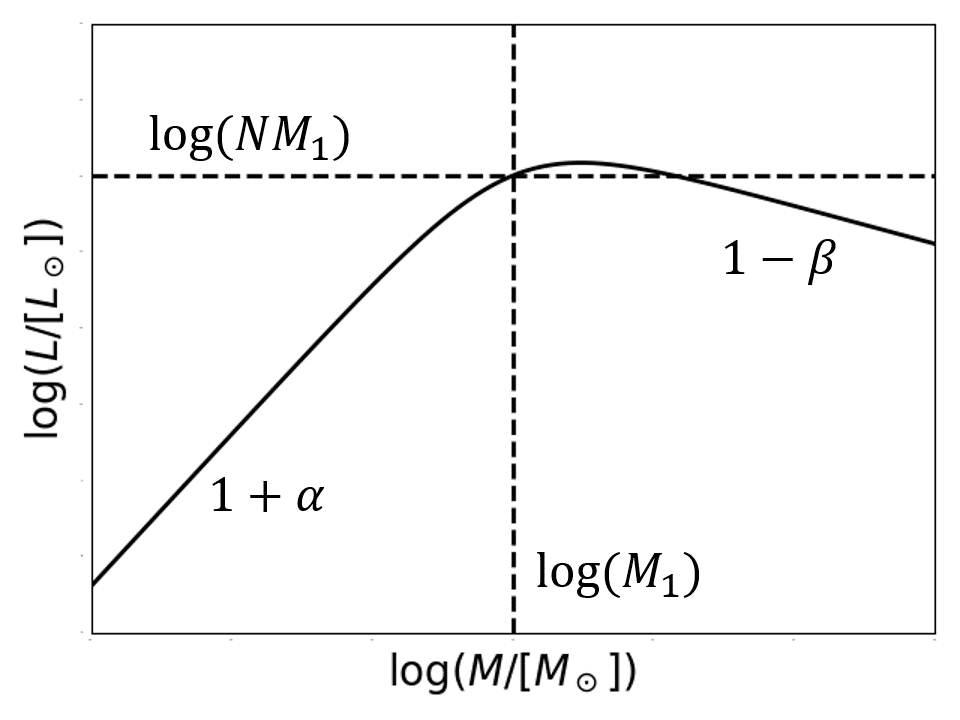}
    \caption{Sketch of the double power law $L(M)$ relation characterized by parameters \{$M_1$, $N$, $\alpha$, $\beta$\} introduced in Eq~\ref{eq:LM}. $M_1$ controls the mass of the turnover of the $L(M)$ relation, $N$ characterizes its amplitude, and $\alpha$ and $\beta$ control the slopes of the $L(M)$ relation at halo mass range $M/[M_\odot]<M_1$ and $M/[M_\odot]>M_1$ respectively. Adapted from \cite{2013MNRAS.428.3121M}.}
    \label{fig:LM_sketch}
\end{figure}

The main observables of LIM surveys, including the intensity of line emission and the two-halo term of the auto-power spectrum, can be estimated through a combination of the average line luminosity versus halo mass $L(M)$ model and a halo model, which provides the number density and bias of halos as a function of their mass and redshift. The $L(M)$ relations and their redshift evolution have been studied via many different approaches in the literature. Although upcoming LIM surveys that cover large cosmic volumes and capture radiation of faint emitters will result in stronger constraints on the $L(M)$ relations, the slopes of $L(M)$ relations at very high and low halo mass are currently poorly determined due to a lack of observational data for very massive halos with $M\gtrsim10^{12}[M_\odot]$ and very faint sources with $M\lesssim10^{11}[M_\odot]$. As a result, current empirical $L(M)$ models in the literature range from linear to power law relations \citep[e.g.][]{2011ApJ...741...70L,2013ApJ...768...15P,2019MNRAS.488.3014P}. In this work we assume a double power law $L(M)$ relation as suggested by \citet{2018MNRAS.475.1477P}:
\begin{equation}\label{eq:LM}
    \dfrac{L}{[L_\odot]}=2N\dfrac{M}{[M_\odot]}\left[\left(\dfrac{M/[M_\odot]}{M_1}\right)^{-\alpha}+\left(\dfrac{M/[M_\odot]}{M_1}\right)^{\beta}\right]^{-1}\,,
\end{equation}
where the four free parameters $M_1$, $N$, $\alpha$, and $\beta$ control the mass of the turnover, amplitude, slope at $M/[M_\odot]<M_1$, and slope at $M/[M_\odot]>M_1$ for the $L(M)$ relation. A schematic diagram of how \{$M_1$, $N$, $\alpha$, $\beta$\} control the shape of the $L(M)$ model is presented in Figure \ref{fig:LM_sketch}. A double power law relation has frequently been used in the literature to describe galaxy-dark matter connections including stellar mass or luminosity versus halo mass relations \citep{2012ApJ...752...41Y,2013MNRAS.428.3121M}. Stellar mass - halo mass empirical models that are consistent with observational constraints (e.g. \cite{2010ApJ...717..379B,2013ApJ...770...57B,2019MNRAS.488.3143B}) indicate that the halo star formation efficiency increases with halo mass from $M\sim10^{10}M_\odot$ to a redshift dependent threshold $M_1$. For more massive halos, the star formation efficiency decreases with halo mass instead, leading to a shallower SFR-M slope. Massive star/supernovae (SNe) feedback drives the slope of the galaxy-halo relation at masses below $M_1$, and active galactic nucleus (AGN) feedback is the main driver at higher halo masses. Specifically, stellar feedback depletes the cold gas reservoir of low mass halos via strong outflows, which are less efficient at removing gas in more massive halos with deeper potential wells. On the other hand, massive halos are more likely to harbor massive bulge-dominated galaxies, which host supermassive black holes. These black holes drive AGN feedback in the form of radiation pressure driven winds, which can remove cold gas from the galaxy, and powerful jets, which can heat the hot gas in the halo and suppress or quench cooling. As a result, the dependence of the efficiency of stellar driven winds on galaxy circular velocity is expected to determine the slope of the double power law at $M<M_1$ ($\alpha$), the efficiency of the jet mode AGN feedback coupling with the hot gas is expected to determine the slope at $M>M_1$, and the relative importance of these two feedback channels determines where they cross, which determines the value of $M_1$. Since the ISM sub-mm lines considered in this work all trace relatively dense star forming gas, the line luminosities are strongly correlated to SFR. It is therefore natural to expect the average $L(M,z)$ relations to follow similar trends.\par 

\begin{figure*}
    \centering
    \includegraphics[width=1\textwidth]{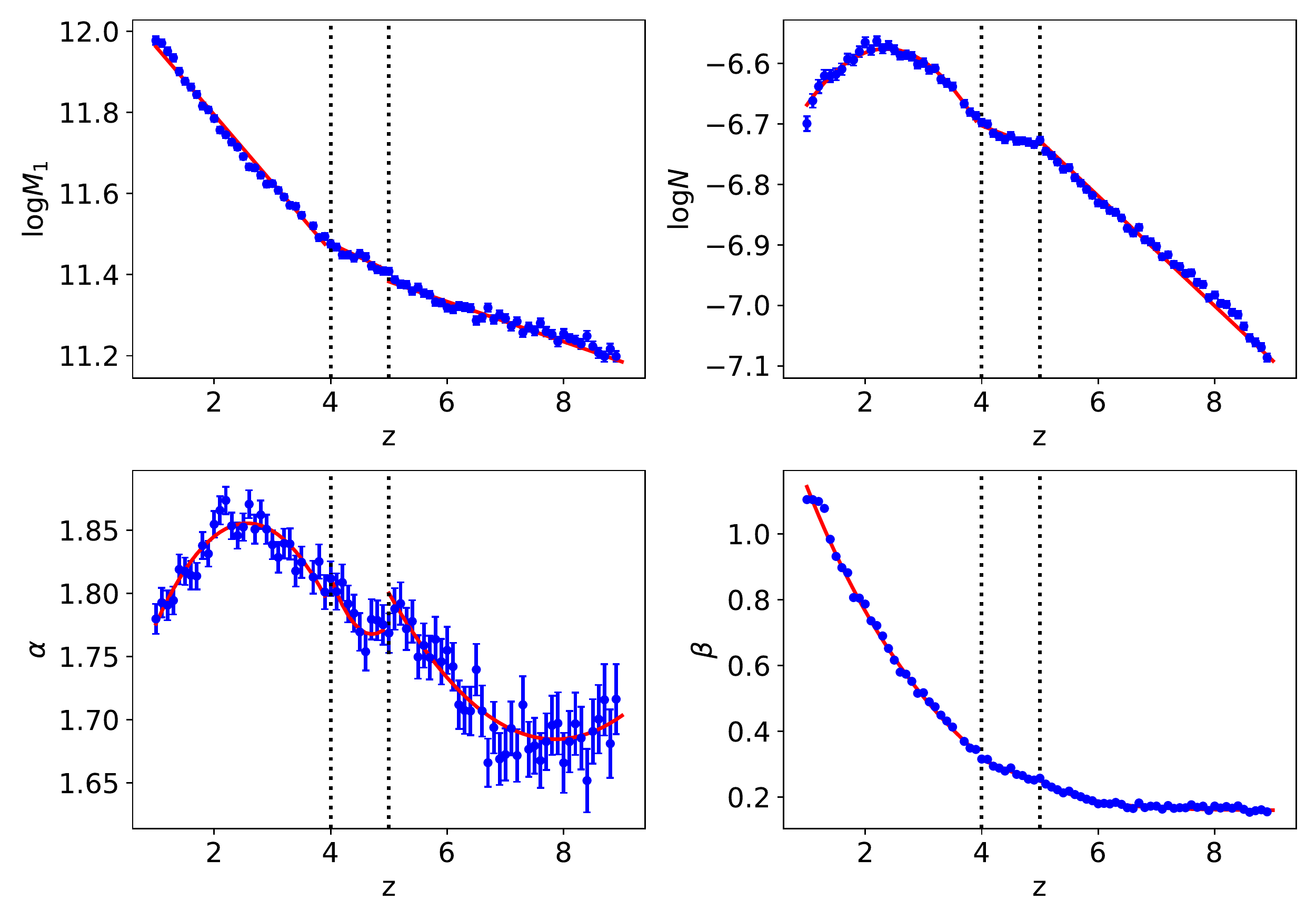}
    \caption{Redshift evolution of parameters \{$M_1$, $N$, $\alpha$, $\beta$\} introduced in Eq~\ref{eq:LM} for CO J=1-0. Blue points show the best-fit parameter values, while the red curves are our models summarized in Table \ref{tb:LM_paramz}.}
    \label{fig:paramz}
\end{figure*}

In each of the 91 redshift slices uniformly distributed in $0\leq z\leq9$, we divide $10^4$ halos simulated by the sub-mm-SAM into 30 mass bins. The sub-mm line luminosity of each halo is determined by the total luminosity of all the galaxies within the halo. We then select star forming halos whose central galaxy satisfies $\mathrm{sSFR}>1/(3t_\mathrm{H}(z))$ to calculate the average $L(M)$ relation. Here sSFR is the specific star formation rate defined as ratio between the galaxy SFR and stellar mass, and $t_\mathrm{H}(z)$ is the Hubble time at the halo redshift. We then fit for the values of the parameters in Eq~\ref{eq:LM} using the tabulated $L(M)$ relations from the sub-mm SAM through a non-linear least square approach. We find the average $L(M)$ statistic predicted by the sub-mm SAM is well represented by the double power law relation. The redshift evolution of the parameters \{$M_1$,$N$,$\alpha$,$\beta$\} is summarized in Table \ref{tb:LM_paramz}. In order to describe the redshift evolution of the double power law model parameters with simple functions, we bin the results into coarser redshift bins $1\leq z<4$, $4\leq z<5$, $5\leq z\leq9$ and model the parameter redshift dependence separately. As one example, in Figure \ref{fig:paramz} we show the comparison between \{$M_1$,$N$,$\alpha$,$\beta$\} best fit values (blue points) and our model summarized in Table \ref{tb:LM_paramz} (red curves) for the CO J=1-0 line. 

We find that in the redshift range $1\leq z\leq9$, the characteristic halo mass $M_1$ that corresponds to the star formation efficiency peak decreases from $10^{12} M_\odot$ to $10^{11}M_\odot$, in agreement with \cite{2013ApJ...770...57B}. The parameter $N$ that controls the overall amplitude of $L(M,z)$ peaks at $z\sim2$, consistent with the observed cosmic star formation rate density redshift evolution. The parameter $\alpha$ that controls the $L(M)$ slope on the $M<M_1$ side shows a relatively weak redshift dependence. This is because the slope of the dependence of the mass loading of stellar driven winds on galaxy circular velocity is assumed to be a fixed value in these models. There is a weak dependence on redshift due to the weakly evolving relationship between halo mass and galaxy circular velocity.  The parameter $\beta$ that characterizes the $L(M)$ relation slope on the high halo mass end increases much more significantly from $z\sim6$ to $z=1$, indicating that on average, the star formation activity within massive halos becomes less efficient as time goes by. This is because in the SC SAM, black hole growth is tied to bulge growth, and massive bulges build up at late times via gas-poor mergers. The radio jet power is assumed to be proportional to the black hole mass, motivated by observations \citep{2008MNRAS.391..481S}. As the black hole mass increases over time, the gas heating rate due to these jets can approach or exceed the cooling rate in massive halos, leading to quenching of star formation and a decrease of the $L(M)$ slope on the high mass end.\par 

\begin{figure*}
    \centering
    \includegraphics[width=1\textwidth]{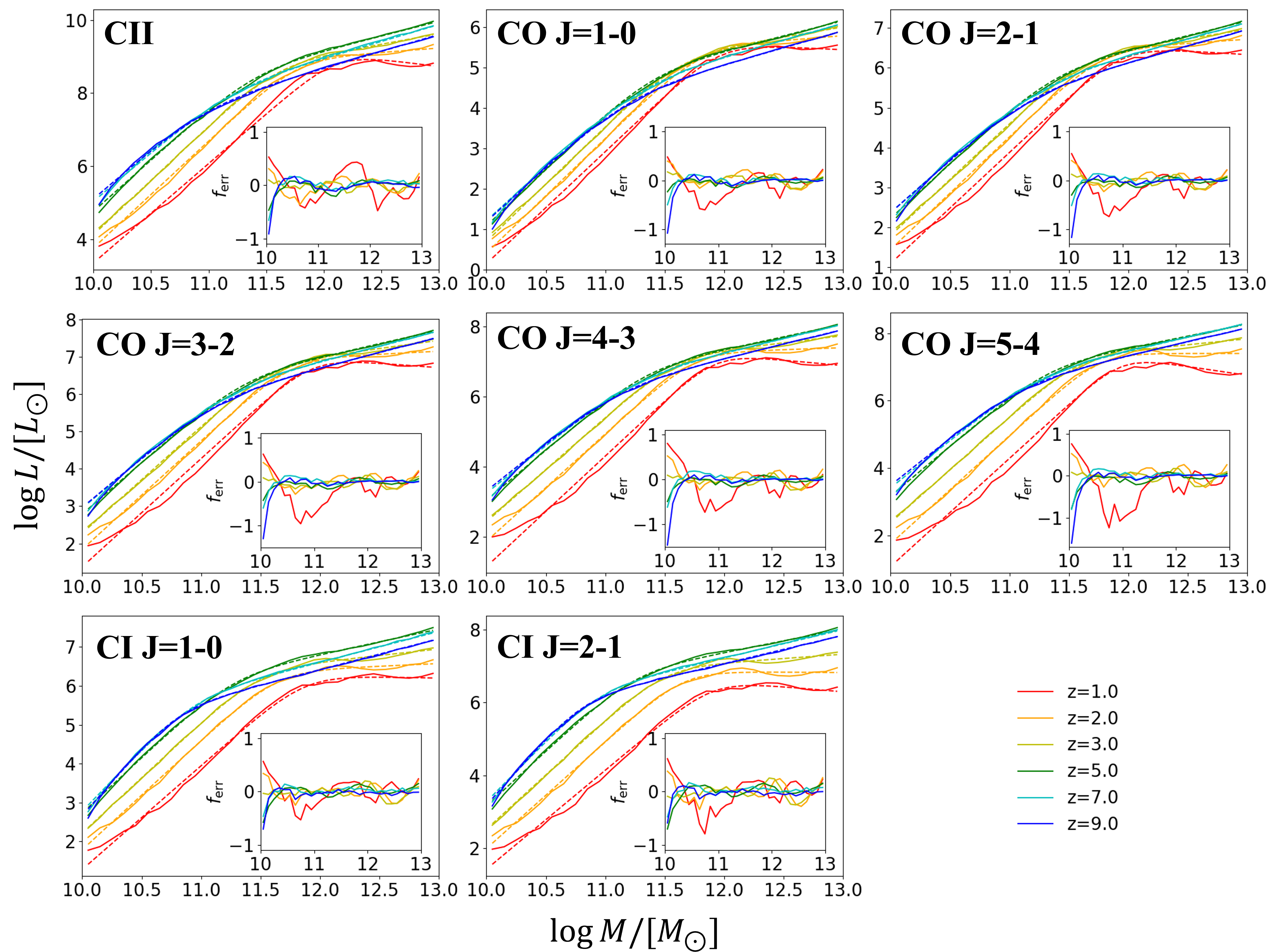}
    \caption{Comparisons of the average value of $L(M)$ between the sub-mm SAM simulation predictions (solid lines) and the double power law empirical model (dashed lines) at $1\leq z\leq9$. The smaller panel in each subplot shows the fractional difference of $L(M)$ relations between sub-mm SAM and the empirical model $f_\mathrm{err}=(L_\mathrm{SAM}-L_\mathrm{model})/L_\mathrm{SAM}$.}
    \label{fig:LM_compare}
\end{figure*}

We present example comparisons between $L(M)$ relations from the sub-mm SAM and our empirical model for [\cii], CO J=1-0 to J=5-4, and [\ci] J=1-0 to J=2-1 lines in Figure \ref{fig:LM_compare}, with fractional differences $(L_\mathrm{SAM}-L_\mathrm{model})/L_\mathrm{SAM}$ specified in the smaller panel of each subplot. Our double power law $L(M,z)$ model generally recovers the sub-mm SAM predictions with better than 20\% accuracy, but the fractional difference can be large at low halo mass. Since halos with $M\leq10^{11}M_\odot$ are very faint sub-mm line emission sources, we will show in section \ref{sec:crosscheck} that the inaccuracy of our $L(M)$ model at $M\leq 10^{11}M_\odot$ does not corrupt its LIM summary statistics predictions.\par
\begin{table*}
\centering
\begin{tabular}{ |l | c | c|c|c|} 
\hline\hline
\multicolumn{5}{c}{$1.0\leq z<4.0$}\\
\hline
line& log$M_1$&log$N$&$\alpha$&$\beta$  \\ 
\hline
\cii &$12.11z^{-0.04105}$&$-0.907\exp(-z/0.867)-3.04$&$1.35+0.450z-0.0805z^2$&$2.57\exp(-z/1.55)+0.0575$\\ 
\hline
CO J=1-0 &$12.13-0.1678z$&$-6.855+0.2366z-0.05013z^2$&$1.642+0.1663z-0.03238z^2$&$1.77\exp(-z/2.72)-0.0827$ \\ 
\hline
CO J=2-1 &$12.12-0.1704z$&$-5.95+0.278z-0.0521z^2$&$1.69+0.126z-0.0280z^2$&$1.80\exp(-z/2.76)-0.0678$ \\
\hline
CO J=3-2 &$12.1-0.171z$&$-5.53+0.329z-0.0570z^2$&$1.843+0.08405z-0.02485z^2$&$1.88\exp(-z/2.74)-0.0623$  \\
\hline
CO J=4-3 &$12.09-0.1662z$&$-5.35+0.404z-0.0657z^2$&$2.24-0.0891z$&$2.017\exp(-z/2.870)-0.1127$ \\
\hline
CO J=5-4 &$12.12-0.1796z$&$-5.37+0.515z-0.0802z^2$&$2.14+0.110z-0.0371z^2$&$2.39\exp(-z/2.55)-0.0890$  \\
\hline
CI J=1-0 &$1.026\exp(-z/2.346)+11.20$&$-1.451\exp(-z/1.893)-5.046$&$-0.741\exp(-z/0.739)+1.86$&$1.26-0.198z$ \\
\hline
CI J=2-1 &$1.135\exp(-z/2.616)+11.13$&$-2.03\exp(-z/1.80)-4.44$&$-1.16\exp(-z/0.706)+2.00$&$1.54-0.259z$  \\
\hline
\multicolumn{5}{c}{$4.0\leq z<5.0$}\\
\hline
\cii &$8.69+1.26z-0.143z^2$&$-5.467+1.056z-0.1133z^2$&$6.135-1.786z+0.1837z^2$&$0.443-0.0521z$ \\ 
\hline
CO J=1-0 &$11.75-0.06833z$&$-6.554-0.03725z$&$3.73-0.833z+0.0884z^2$&$0.598-0.0710z$  \\ 
\hline
CO J=2-1 &$11.74-0.07050z$&$-5.57-0.0250z$&$4.557-1.215z+0.1300z^2$&$0.657-0.0794z$ \\
\hline
CO J=3-2 &$11.74-0.07228z$&$-5.06-0.0150z$&$5.253-1.502z+0.1609z^2$&$0.707-0.0879z$  \\
\hline
CO J=4-3 &$11.73-0.07457z$&$-4.784$&$5.74-1.67z+0.178z^2$&$0.762-0.0984z$  \\
\hline
CO J=5-4 &$11.73-0.07798z$&$-3.81-0.359z+0.0419z^2$&$6.12-1.77z+0.188z^2$&$0.846-0.115z$ \\
\hline
CI J=1-0 &$8.54+1.33z-0.157z^2$&$-6.96+0.750z-0.0807z^2$&$8.42-2.91z+0.320z^2$&$0.837-0.103z$\\
\hline
CI J=2-1 &$8.823+1.206z-0.1445z^2$&$-4.906+0.05632z$&$9.275-3.225z+0.3543z^2$&$0.94-0.12z$  \\
\hline
\multicolumn{5}{c}{$z\geq5$}\\
\hline
\cii &$11.92-0.1068z$&$-2.37-0.130z$&$2.82-0.298z+0.0196z^2$&$1760\exp(-z/0.520)+0.0782$ \\ 
\hline
CO J=1-0 &$11.63-0.04943z$&$-6.274-0.09087z$&$2.56-0.223z+0.0142z^2$&$33.4\exp(-z/0.846)+0.160$ \\ 
\hline
CO J=2-1 &$11.63-0.05266z$&$-5.26-0.0849z$&$2.47-0.210z+0.0132z^2$&$38.3\exp(-z/0.841)+0.169$ \\
\hline
CO J=3-2 &$11.62-0.0550z$&$-4.72-0.0808z$&$2.53-0.220z+0.0139z^2$&$31.5\exp(-z/0.879)+0.170$  \\
\hline
CO J=4-3 &$11.6-0.0529z$&$-4.40-0.0744z$&$2.59-0.206z+0.0120z^2$&$41.6\exp(-z/0.843)+0.172$  \\
\hline
CO J=5-4 &$11.58-0.05359z$&$-4.21-0.0674z$&$2.87-0.257z+0.0157z^2$&$21.8\exp(-z/0.957)+0.168$ \\
\hline
CI J=1-0 &$4.294\exp(-z/2.479)+10.68$&$1.89\exp(-z/5.73)-6.03$&$1.04+0.165z$&$2090\exp(-z/0.520)+0.204$  \\
\hline
CI J=2-1 &$4.37\exp(-z/2.38)+10.7$&$1.48\exp(-z/6.14)-5.31$&$1.17+0.169z$&$3960\exp(-z/0.487)+0.221$  \\
\hline
\end{tabular}
\caption{Redshift evolution of parameters in the $L(M)$ double power law model Eq~\ref{eq:LM}.}
\label{tb:LM_paramz}
\end{table*}

\subsection{Halo duty cycle factor}\label{sec:fduty}
\begin{figure}
    \centering
    \includegraphics[width=0.49\textwidth]{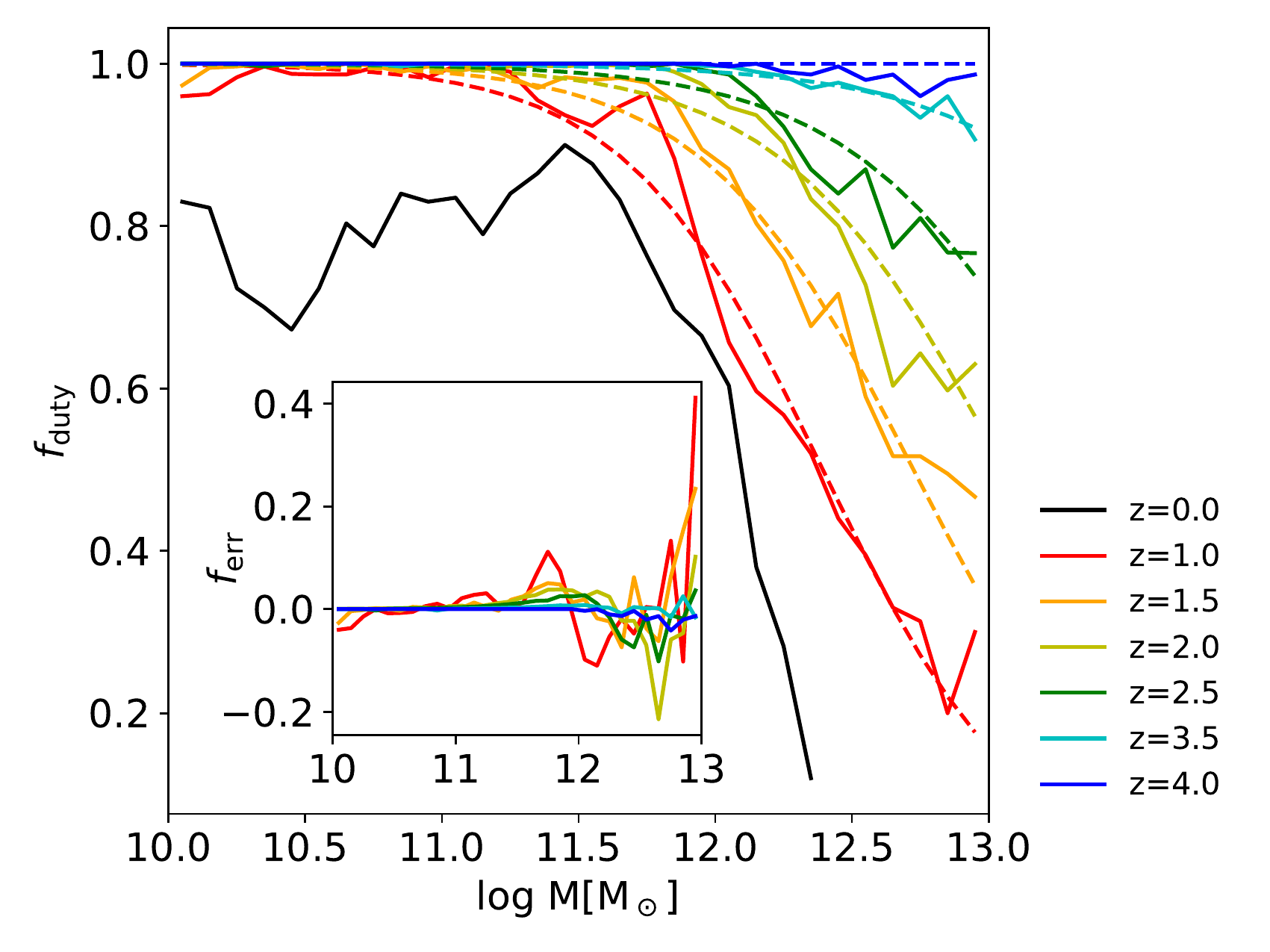}
    \caption{$f_\mathrm{duty}(M)$ comparisons between SAM simulations and the double power law empirical model predictions at $1\leq z\leq4$. The solid curves show $L(M)$ relations simulated by SAM $+$ sub-mm SAM, while the dashed lines are given by the empirical model Eq~\ref{eq:fduty}. The smaller panel shows the fractional difference of $f_\mathrm{duty}(M)$ relations between SAM and the empirical model $f_\mathrm{err}=(f_\mathrm{duty}^\mathrm{SAM}-f_\mathrm{duty}^\mathrm{model})/f_\mathrm{duty}^\mathrm{SAM}$. At $z<1$, the SAM $f_\mathrm{duty}(M)$ relations become more complicated than a double power law due to low mass halo quenching by stellar feedback, as shown by the black curve.}
    \label{fig:fduty_compare}
\end{figure}

As mentioned in section~\ref{sec:LM}, we have removed quenched dark matter halos that are not forming stars rapidly and therefore are not likely to be luminous in [\cii], CO, and [\ci] lines from the $L(M)$ model fitting, so another important part of our empirical treatment is to model the fraction of star forming halos $f_\mathrm{duty}$ in each halo mass bin and redshift slice. We calculate $f_\mathrm{duty}$ as the fraction of halos in each mass and redshift bin with a central galaxy that satisfies the condition $\mathrm{sSFR}>1/(3t_\mathrm{H}(z))$, as introduced above:
\begin{equation}
    f_\mathrm{duty}(M,z)=\dfrac{N_\mathrm{SF}(M,z)}{N_\mathrm{total}(M,z)}\,,
\end{equation}
here $N_\mathrm{SF}$ is the number of star-forming halos, while $N_\mathrm{total}$ is the total number of dark matter halos in the corresponding redshift and halo mass bin. We find the $f_\mathrm{duty}(M)$ relation can be described by a double power law at redshift $1\leq z<4$:
\begin{equation}\label{eq:fduty}
    f_\mathrm{duty}= 
\begin{cases}
    \dfrac{1}{1+\left(\frac{M/[M_\odot]}{M_2}\right)^\gamma},& \text{if } 1\leq z< 4\\
    1,              & \text{if } 4\leq z\leq9
\end{cases}
\end{equation}
with redshift evolution given by:
\begin{equation}
    \begin{split}
        \log M_2(z)&=11.73+0.6634z\,,\\
        \gamma(z)&=1.37-0.190z+0.0215z^2\,.
    \end{split}
\end{equation}
The gas accretion rate into halos decreases at $z<1$ due to the onset of the accelerating cosmic expansion. As a result, central galaxies living in less massive halos ($M\lesssim10^{11}M_\odot$) start to become quenched by stellar feedback, causing a more complicated $f_\mathrm{duty}(M)$ behavior at low halo mass that cannot be fully captured by this simple double power law relation. We show example comparisons between the $f_\mathrm{duty}(M)$ relations predicted by the sub-mm SAM and this model in Figure \ref{fig:fduty_compare}, with fractional difference $f_\mathrm{err}=(f_\mathrm{duty}^\mathrm{SAM}-f_\mathrm{duty}^\mathrm{model})/f_\mathrm{duty}^\mathrm{SAM}$ specified in the smaller panel. The accuracy of our $f_\mathrm{duty}$ model is better than 20\% in redshift range $1\leq z\leq 9$ and halo mass range $10^{10}M_\odot\leq M\leq 10^{12.5}M_\odot$.
The strong redshift dependence of $f_\mathrm{duty}$ at high halo masses is driven by the growth of massive black holes and the corresponding increase in the efficiency of AGN feedback, as described earlier. 

\subsection{$L(M)$ scatter and multi-line correlations}\label{sec:sigma}
With the average line luminosity versus halo mass $L(M)$ model and the halo duty cycle factor $f_\mathrm{duty}(M)$ model introduced in section \ref{sec:LM} and section \ref{sec:fduty}, key LIM observables can be calculated quantitatively by combining these with halo models (e.g. \cite{2002PhR...372....1C,2021arXiv210301964S}). The intensities of [\cii], CO, and [\ci] lines can be estimated through:
\begin{equation}\label{eq:intensity}
    I_i(z)=\dfrac{c}{4\pi\nu^0_iH(z)}\int dM\dfrac{dn}{dM}L_i(M,z)f_\mathrm{duty}\,,
\end{equation}
here $i$ is an index for sub-mm lines considered in this work. $\nu^0_i$ is the rest frame frequency of the target emission line. $dn/dM$ is the halo mass function. In this work we choose the \cite{2002MNRAS.329...61S} halo mass function model since it matches well with halo abundances at high redshift predicted by hydrogen reionization simulations \citep{2007ApJ...654...12Z}. However, the $L(M)$ model introduced in this work can also be combined with other halo mass function models that are calibrated to simulations and observations at lower redshifts. Two-halo terms of the auto/cross power spectrum can also be estimated through:
\begin{equation}\label{eq:P2-halo}
    P^\mathrm{2-halo}_{ij}(k,z)=I_i(z)I_j(z)\langle b_i(z)\rangle\langle b_j(z)\rangle P_\mathrm{lin}(k,z)\,,
\end{equation}
Here $P_\mathrm{lin}$ is the linear theory density power spectrum, $i$ and $j$ are indices for sub-mm lines considered in this work, and $\langle b\rangle$ is the luminosity weighted halo bias:
\begin{equation}
    \langle b_i(z)\rangle=\dfrac{\int dM\frac{dn}{dM}f_\mathrm{duty}L_i(M,z)b(M,z)}{\int dM\frac{dn}{dM}f_\mathrm{duty}L_i(M,z)}\,.
\end{equation}
In this work we use the halo bias model $b(M)$ proposed by \cite{2001MNRAS.323....1S}. At small scales where Poisson noise dominates the power spectrum term, models for the scatter and correlations of $L(M)$ relations become important for the shot noise power spectrum estimation:
\begin{equation}\label{eq:Pshot}
\begin{split}
    P^\mathrm{shot}_{ij}(z)&=\left(\dfrac{c}{4\pi H(z)}\right)^2\dfrac{1}{\nu_i^0\nu_j^0}\times\\
    &\int dM\dfrac{dn}{dM}\langle L_i(M,z)L_j(M,z)\rangle f_\mathrm{duty}(M,z)\,,
\end{split}
\end{equation}
here $\langle L_i(M,z)L_j(M,z)\rangle$ denotes the average line $i$ and line $j$ luminosity product of all star-forming halos at a given halo mass at redshift $z$.\par
We find that in the sub-mm SAM simulations, the luminosity distribution of star-forming halos within a narrow halo mass bin is close to log-normal with a mean given by Eq~\ref{eq:LM} and a dispersion of $\sigma$ dex:
\begin{equation}
\begin{split}
    &\dfrac{dP}{d\log \hat{L}}=\\
    &\dfrac{1}{\sigma\sqrt{2\pi}}\exp\left[-\dfrac{1}{2}\left(\dfrac{\log \hat{L}-\log L+\ln(10)\sigma^2/2}{\sigma}\right)^2\right]\,.
\end{split}
\end{equation}
Here we use $\hat{L}$ to denote the halo luminosities predicted by the sub-mm SAM. $L$ denotes the average halo luminosity given by the model presented in section \ref{sec:LM}. Note that this approximation only holds if we remove quenched galaxies from our distributions, necessitating the inclusion of the above $f_{\rm{duty}}$ factor.  This differs from previous works which tend to include one or the other of a scattering or duty cycle effect \citep[see, e.g.][]{Keating2016}.

The average line luminosity product under this assumption on the $L(M)$ distribution is given by:
\begin{equation}\label{eq:cov}
    \begin{split}
        &\langle L_i(M,z)L_j(M,z)\rangle=\\
        &(\rho_{ij}\sqrt{10^{\ln(10)\sigma_i^2}-1}\sqrt{10^{\ln(10)\sigma_j^2}-1}+1)L_i(M,z)L_j(M,z)\,,
    \end{split}
\end{equation}
here $\rho_{ij}$ is the correlation coefficient between $L_i(M,z)$ and $L_j(M,z)$. Consider the $i=j$ case where we study the variance of one emission line. In this case, $\rho$ will always be unity and Eq~\ref{eq:cov} reduces to:
\begin{equation}
    \langle L_i^2(M,z)\rangle=10^{\ln(10)\sigma_i^2}L_i^2(M,z)\,.
\end{equation}
We ignore the halo mass dependence of the dispersion in $L(M)$. Instead of constructing a comprehensive $\sigma(M,z)$ empirical model for the dispersion, we equate the shot noise power spectrum simulated by the sub-mm SAM to the empirical model prediction and solve for a characteristic $L(M)$ scatter. We summarize the redshift evolution of $\sigma$ for each sub-mm emission line in Eq \ref{eq:sigma}.\par
\begin{equation}\label{eq:sigma}
    \begin{split}
        \sigma_\mathrm{CII}(z)&=0.32\exp(-z/1.5)+0.18\,,\\
        \sigma_\mathrm{CO J=1-0}(z)&=0.357-0.0701z+0.00621z^2\,,\\
        \sigma_\mathrm{CO J=2-1}(z)&=0.36-0.072z+0.0064z^2\,,\\
        \sigma_\mathrm{CO J=3-2}(z)&=0.40-0.083z+0.0070z^2\,,\\
        \sigma_\mathrm{CO J=4-3}(z)&=0.42-0.091z+0.0079z^2\,,\\
        \sigma_\mathrm{CO J=5-4}(z)&=0.44-0.085z+0.0063z^2\,,\\
        \sigma_\mathrm{CI J=1-0}(z)&=0.39-0.076z+0.0063z^2\,,\\
        \sigma_\mathrm{CI J=2-1}(z)&=0.46-0.096z+0.0079z^2\,.
    \end{split}
\end{equation}\par
\begin{figure}
    \centering
    \includegraphics[width=0.45\textwidth]{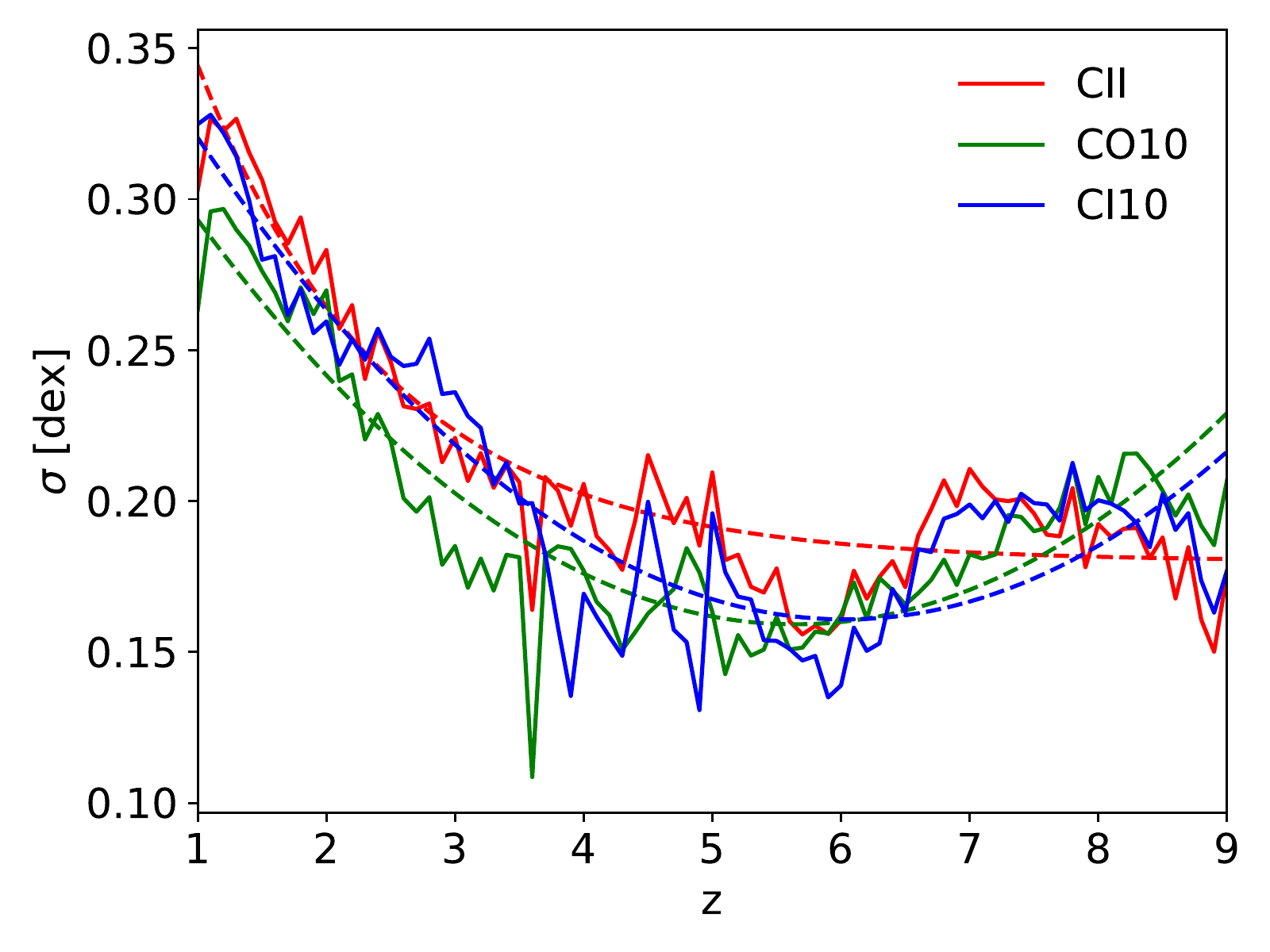}
    \caption{Comparisons of the line emission scatter $\sigma(z)$ between the sub-mm SAM simulation predictions (solid lines) and the model given by Eq \ref{eq:sigma} (dashed lines) at $1\leq z\leq9$ for [\cii] (red), CO J=1-0 (green), and [\ci] J=1-0 (blue). }
    \label{fig:sigma}
\end{figure}

We present the $\sigma(z)$ comparisons between sub-mm SAM and Eq \ref{eq:sigma} for [\cii], CO J=1-0, and [\ci] J=1-0 in Figure \ref{fig:sigma}.

\begin{figure*}
    \centering
    \includegraphics[width=0.5\textwidth]{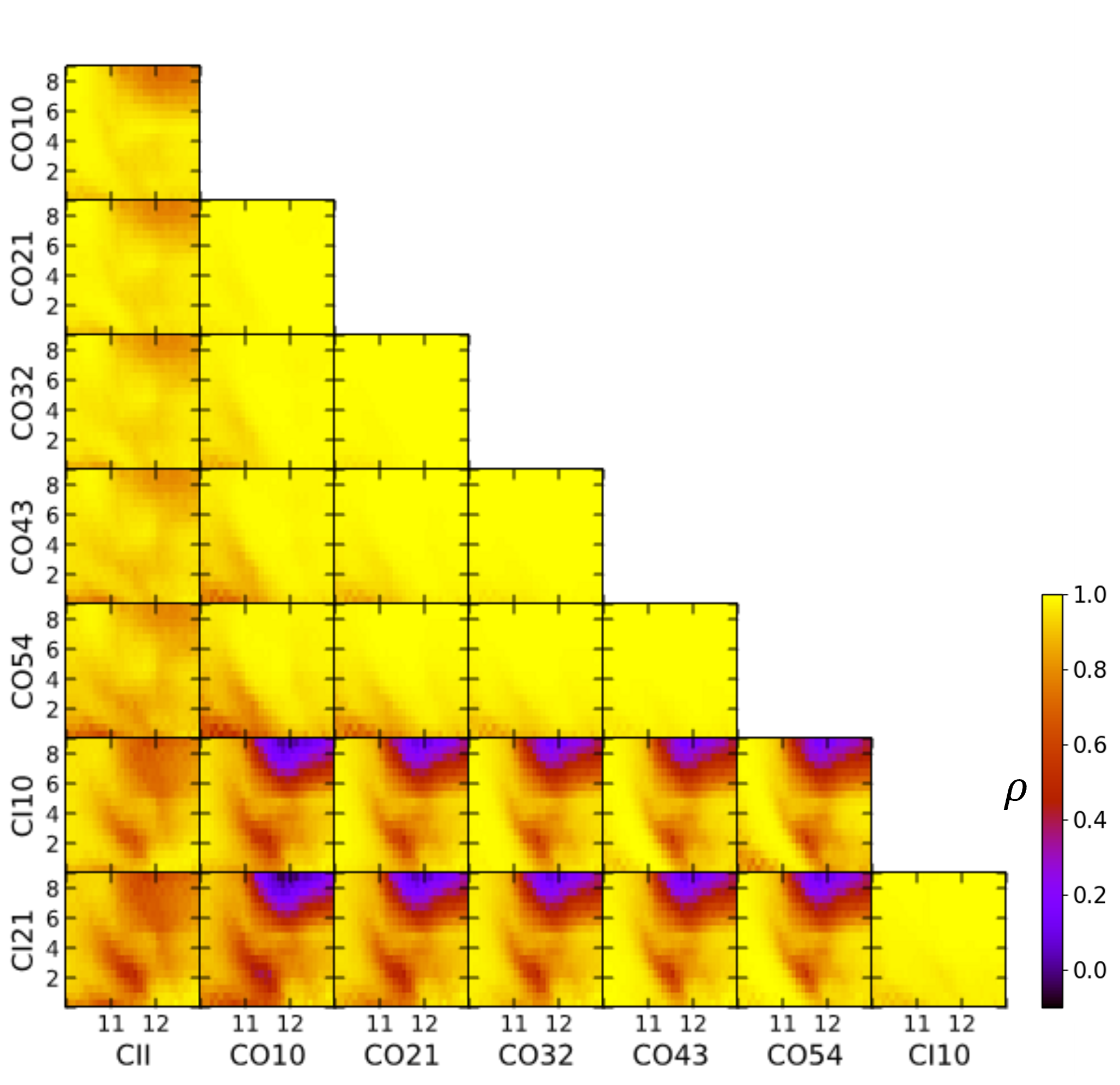}
    \includegraphics[width=0.49\textwidth]{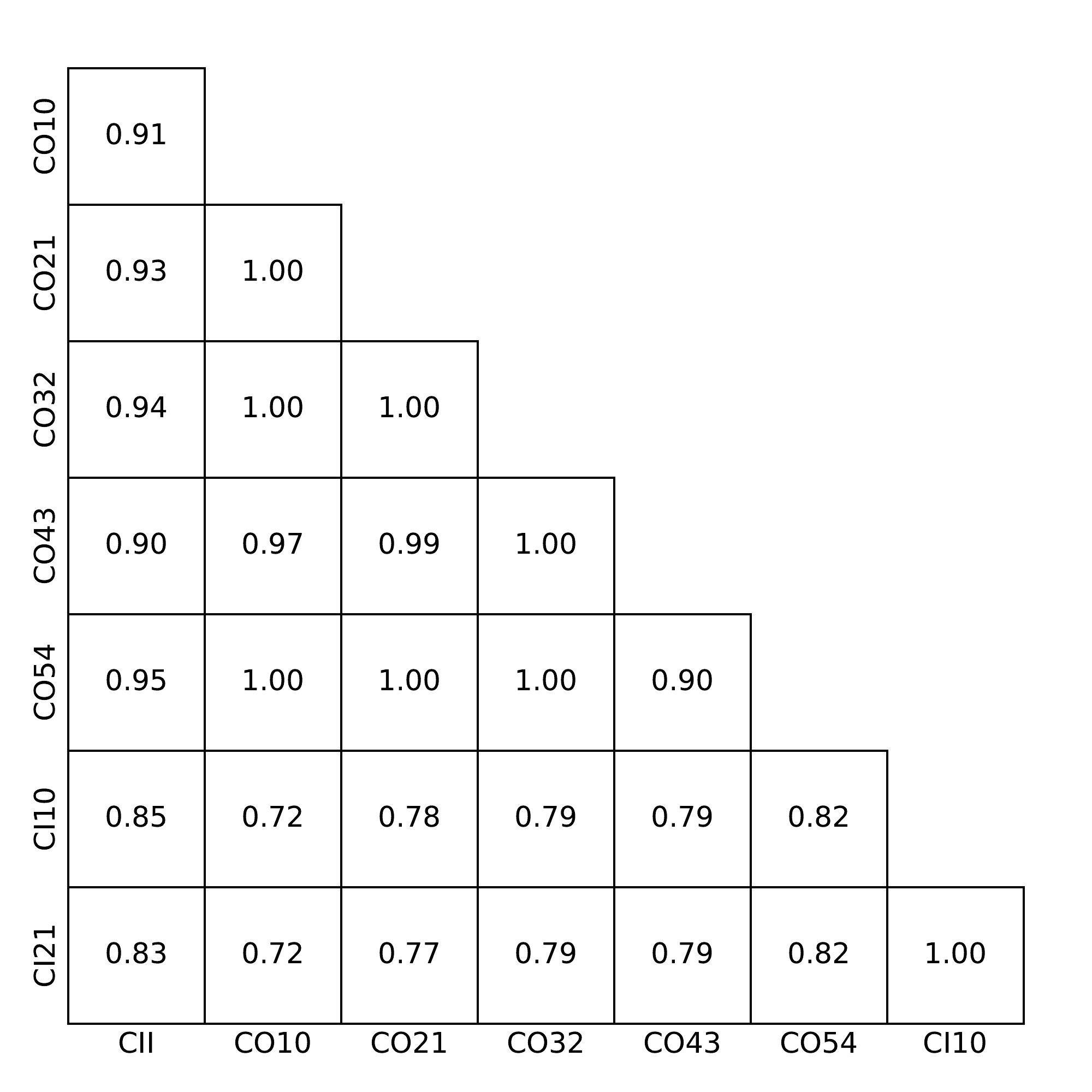}
    \caption{Cross-correlation coefficients $\rho$ among sub-mm emission lines considered in this work. The outer x and y axis labels specify the target emission lines. \textit{Left:} The inner x axis of each cell shows the halo mass $10\leq\log(M/[M_\odot])\leq13$, divided into 30 log mass bins. The y axis shows the redshift of the sub-mm SAM simulation output $0\leq z\leq9$, divided into 18 bins. The color map specifies $\rho$ of star-forming halos in the corresponding halo mass and redshift bin. Besides the CO cross [\ci] lines at $z\gtrsim6$ and $\log(M/[M_\odot])\gtrsim11$, we find the cross-correlation coefficients show weak mass and redshift dependence. \textit{Right:} The [\cii], CO, and [\ci] line cross-correlation coefficients of all the star-forming halos in the sub-mm SAM.}
    \label{fig:rho}
\end{figure*}

Assuming a log-normal $L(M)$ distribution, another quantity we need to model is the correlation coefficient $\rho$. Through studying the correlation among [\cii], CO, and [\ci] lines simulated by the sub-mm SAM, we find the correlation coefficient generally shows weak dependence over $\log (M/[M_\odot])\lesssim11$ and $z\lesssim4$, as presented in Figure \ref{fig:rho} left panel. The CO-[\ci] line correlation weakens  at high redshift and in the high halo mass range for the following reason: the molecular gas and stellar mass surface density of individual galaxies in this part of parameter space are high due to rapid gas accretion and relatively weak stellar and AGN feedback. Since the sub-mm SAM assumes the UV radiation field strength to be proportional to the star formation rate surface density, molecular clouds in halos with $\log (M/[M_\odot])>11$ and $z>4$ are subjected to strong UV radiation fields. As a result, a large amount of carbon is singly ionized, leaving limited volume where \ci\ and CO compete with each other. We define the characteristic correlation coefficient $\rho_{ij}$ between two target lines as a value that equates the shot noise power spectrum given by the SAM and the empirical model:
\begin{equation}
\begin{split}
    &\int d\ln M\dfrac{dn}{d\ln M}\langle L_i(M,z)L_j(M,z)\rangle f_\mathrm{duty}\\
    =&(\rho_{ij}(z)\sqrt{10^{\ln(10)\sigma_i^2}-1}\sqrt{10^{\ln(10)\sigma_j^2}-1}+1)\\
    \times&\int d\ln M\dfrac{dn}{d\ln M}L_i(M,z)L_j(M,z)f_\mathrm{duty}
\end{split}
\end{equation}
The resulting cross-correlation $\rho$ factors averaged over all redshift snapshots are presented in Figure \ref{fig:rho} right panel. The characteristic correlation coefficients simulated by the sub-mm SAM are mostly consistent with the empirical galaxy generator (\textsc{EGG}) simulation results \citep{2021arXiv210301964S}. Despite adopting independent simulation approaches, we both find the correlation coefficients among CO and among [\ci] lines are close to unity. This is a natural prediction because CO lines are all emitted by the CO molecules. The luminosity ratios among CO lines are therefore less sensitive to gas phase metallicity, external radiation field, and many other environmental variables. The same argument is also applicable to [\ci] lines. However, we find the $\rho_\mathrm{CO, CI}$ factors predicted by the sub-mm SAM are slightly lower compared to the predictions of the \text{EGG} simulator, while $\rho_\mathrm{CII,CO}$ and $\rho_\mathrm{CII,CI}$ factors predicted by the sub-mm SAM are significantly higher than \text{EGG}'s predictions. The $\rho_\mathrm{CO, CI}$ difference could be caused by the fact that the EGG correlation coefficients are only studied out to redshift $z=4$, while our work extends to $z=9$. Note that both the ionized and neutral ISM phases contribute to the [\cii] line radiation. The difference among $\rho_\mathrm{CII,CO}$ and $\rho_\mathrm{CII,CI}$ could be caused by our different assumptions about the molecular cloud masses. If the molecular clouds are assumed to be small, the galaxy star-formation activity can ionize the whole cloud such that it becomes [\cii] luminous while very faint in CO and [\ci] lines. This will lead to small or even negative correlations among [\cii], CO, and [\ci] lines. On the other hand, if the molecular clouds in a simulation are large enough such that an increase of HII region volume will not significantly influence the molecular gas reservoir, increasing the SFR will boost the strength of all the sub-mm lines considered in this work. In this case we expect [\cii], CO, and [\ci] lines to be strongly positively correlated, as suggested by the sub-mm SAM. This illustrates the sensitively of these kinds of predictions to the uncertain details of the assumed sub-grid model. 

\begin{figure}
    \centering
    \includegraphics[width=0.45\textwidth]{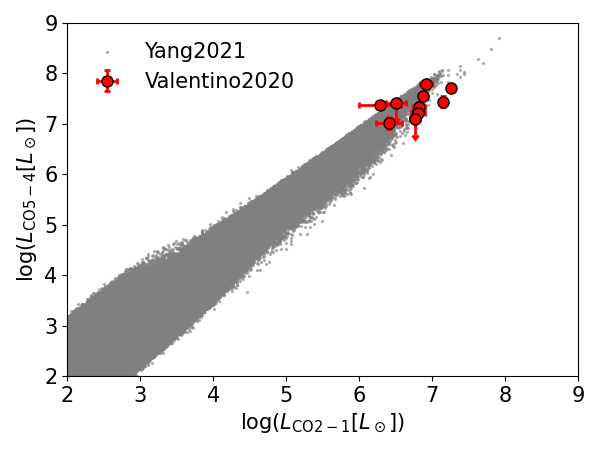}\\
    \includegraphics[width=0.45\textwidth]{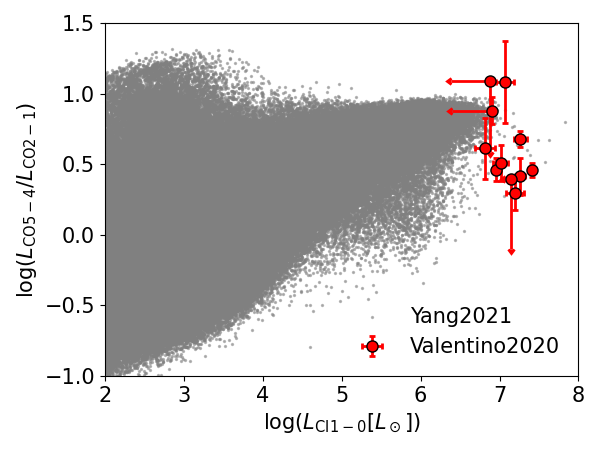}
    \caption{CO SLED comparisons between sub-mm SAM simulations and observations at $1.1\leq z\leq1.3$. Grey points are sub-mm line luminosities of star forming galaxies predicted by the sub-mm SAM adopted in this work combined with a 2 deg$^2$ mock lightcone introduced in \cite{2021ApJ...911..132Y}. Red data points are from Table 2 of \cite{2020ApJ...890...24V}. \textit{Top:} CO J=5-4 versus CO J=2-1 luminosity relation. \textit{Bottom:} Luminosity ratios between CO J=5-4 and CO J=2-1 line versus [\ci] J=1-0 luminosity.}
    \label{fig:COSLED}
\end{figure}

We also compare the $L_\mathrm{CO J=5-4}$ versus $L_\mathrm{CO J=2-1}$ relation and $L_\mathrm{CO J=5-4}/L_\mathrm{CO J=2-1}$ ratios predicted by the SAM with observational measurements as a test of the robustness of the correlation predictions. Specifically, we combine the SAM $+$ sub-mm SAM adopted in this work with the 2 deg$^2$ mock lightcone introduced in \cite{2021ApJ...911..132Y} to simulate sub-mm line luminosities of star-forming galaxies. We then compare the CO J=5-4 versus CO J=2-1 spectral line energy distribution (SLED) predicted by the SAM with observations at $z\sim2$ \citep{2020ApJ...890...24V}. We find that the SAM $+$ sub-mm SAM CO SLED predictions at high [\ci] J=1-0 luminosity are in reasonable agreement with the observed $L_\mathrm{CO J=5-4}/L_\mathrm{CO J=2-1}$ scatter, as shown in Figure \ref{fig:COSLED}. Some of the scatter in the observed relation may of course be due to observational errors. 

\section{Summary statistic cross-check between sub-mm SAM and the empirical model}\label{sec:crosscheck}
In this section we cross-check the LIM statistics predicted by the sub-mm SAM and the empirical model introduced in section \ref{sec:model} in order to test how accurately our model represents the SAM simulation results, as judged by commonly used summary statistics.\par
\begin{figure}
    \centering
    \includegraphics[width=0.49\textwidth]{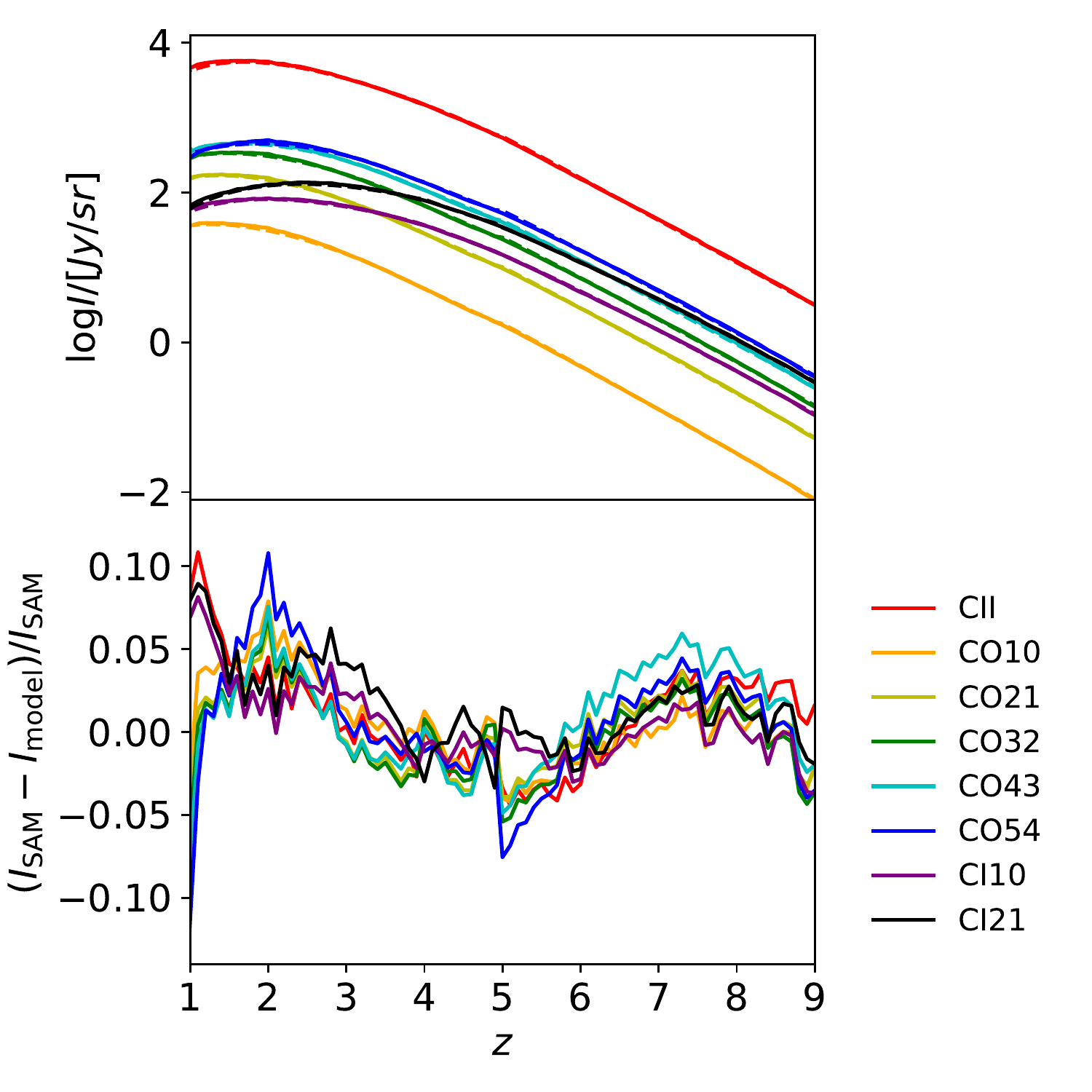}
    \caption{Integrated line intensity comparisons between sub-mm SAM simulations and the empirical model predictions at $1\leq z\leq9$, for various lines as shown in the plot label. \textit{Top}: The solid curves show line intensities from the sub-mm SAM, while dashed lines are obtained from the empirical models. \textit{Bottom}: The fractional differences between sub-mm SAM and empirical model predictions for the integrated line intensity.}
    \label{fig:I_compare}
\end{figure}

We first compare the line intensity predictions. Specifically, we combine the $L(M,z)$ and $f_\mathrm{duty}(M,z)$ relations given by the sub-mm SAM simulation with the halo model Eq~\ref{eq:intensity} to calculate the sub-mm line intensity predictions. We compute the line intensity predictions of the empirical model in a similar way, using the $L(M,z)$ and $f_\mathrm{duty}(M,z)$ relations from Eq~\ref{eq:LM} and Eq~\ref{eq:fduty} instead of the sub-mm SAM predictions for these quantities. We present the line intensity comparisons in Figure \ref{fig:I_compare}. In the redshift range $1\leq z\leq9$, our empirical model predictions agree with the sub-mm SAM simulations to generally better than 10\% accuracy. Fractional differences between the sub-mm SAM and this model's predictions jump at $z=4$ and $z=5$, as shown in the bottom panel of Figure \ref{fig:I_compare}. This is caused by the fact that we fit for parameters in this empirical $L(M,z)$ model in redshift intervals $1\leq z<4$, $4\leq z<5$, and $5\leq z\leq9$ (See Table \ref{tb:LM_paramz}). i.e. the $L(M,z)$ model parameters behave discontinuously at $z=4$ and $z=5$.\par
\begin{figure*}
    \centering
    \includegraphics[width=1\textwidth]{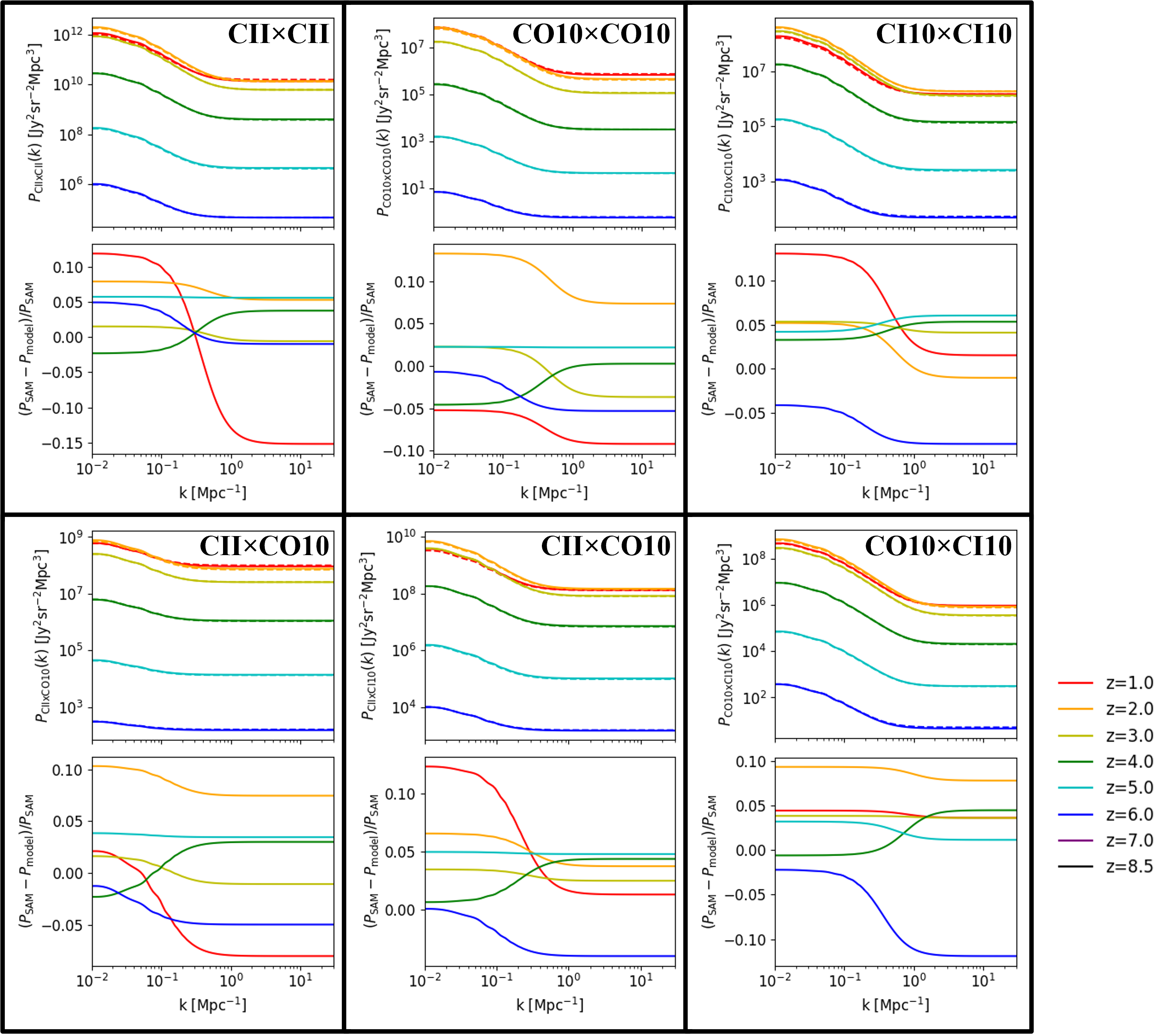}
    \caption{Power spectrum comparisons between sub-mm SAM simulations and the empirical model predictions at $1\leq z\leq9$ among [\cii], CO J=1-0, and [\ci] J=1-0 lines. The first row shows auto LIM power spectra while the bottom row shows the cross power spectra comparisons. In each figure top panel, the solid curves are given by the sub-mm SAM, while dashed lines are predictions of the empirical models. The bottom panel shows the power spectrum fractional difference between sub-mm SAM and the empirical model.}
    \label{fig:P_compare}
\end{figure*}

We then check the line emission auto and cross power spectra. As mentioned in section \ref{sec:model}, the LIM analyses in this work are all calculated at scales larger than those of dark matter halos, so we will ignore the one-halo term in the power spectrum. For simplicity we also ignore the redshift space distortion. We calculate the power spectrum as $P_{ij}(k,z)=P_{ij}^\mathrm{2-halo}(k,z)+P_{ij}^\mathrm{shot}(z)$, where the two-halo term and the shot noise term are given by Eq~\ref{eq:P2-halo} and Eq~\ref{eq:Pshot}. We use $L(M,z)$, $f_\mathrm{duty}$, and $\langle L_i(M,z)L_j(M,z)\rangle$ given by the sub-mm SAM for the simulation power spectra calculations. For the empirical model power spectrum estimation we use the $L(M,z)$, $f_\mathrm{duty}$, $L(M)$ scatter, and correlation coefficient models introduced in section \ref{sec:model}. In Figure \ref{fig:P_compare} we show example comparisons between the sub-mm SAM and the empirical model for the auto and cross power spectra among the [\cii], CO J=1-0, and [\ci] J=1-0 lines. We find the empirical model power spectrum predictions are in agreement with the sub-mm SAM simulations with fractional error less than 25\% at $z\geq1$ The SAM-model agreement increases to better than 15\% at $z\geq 3$.\par

To further test our approach, we also compare with the predictions of a 2 deg$^2$ mock lightcone where halos have been extracted from a dissipationless N-body simulation and filled with galaxies using the sub-mm SAM approach, as described by \citet{2021ApJ...911..132Y}. In order to assess the differences between the modeling approaches relative to the precision of upcoming LIM experiments, we use as a fiducial example a COMAP-like survey with parameters taken from \cite{2016ApJ...817..169L}. Specifically, we create LIM mock data in the frequency range $30\leq\nu/[\mathrm{GHz}]\leq34$ with frequency resolution $\Delta\nu=40\,\mathrm{MHz}$ and angular resolution $\theta_\mathrm{pix}=0.6\,\mathrm{arcmin}$. We then smooth the mock data with a Gaussian beam with full width half maximum (FWHM) $\theta_\mathrm{FWHM}=6\,\mathrm{arcmin}$ such that the smoothed map spatial resolution matches that of COMAP. We assume a map sensitivity 41.5 $\mu$K MHz$^{1/2}$ and estimate the COMAP instrumental noise as a Gaussian probability density distribution (PDF) with zero mean and standard deviation $\sigma_N$, where $\sigma_N$ is calculated as the map sensitivity multiplied by $10/\sqrt{8\ln(2)\Delta\nu}$. Due to the limited cosmic volume and sensitivity of this mock data, we only compare power spectra predicted by different models in the range $0.1\leq k/[\mathrm{Mpc}^{-1}]\leq0.3$. On smaller scales the mock is dominated by the instrumental noise.\par
\begin{figure}
    \centering
    \includegraphics[width=0.49\textwidth]{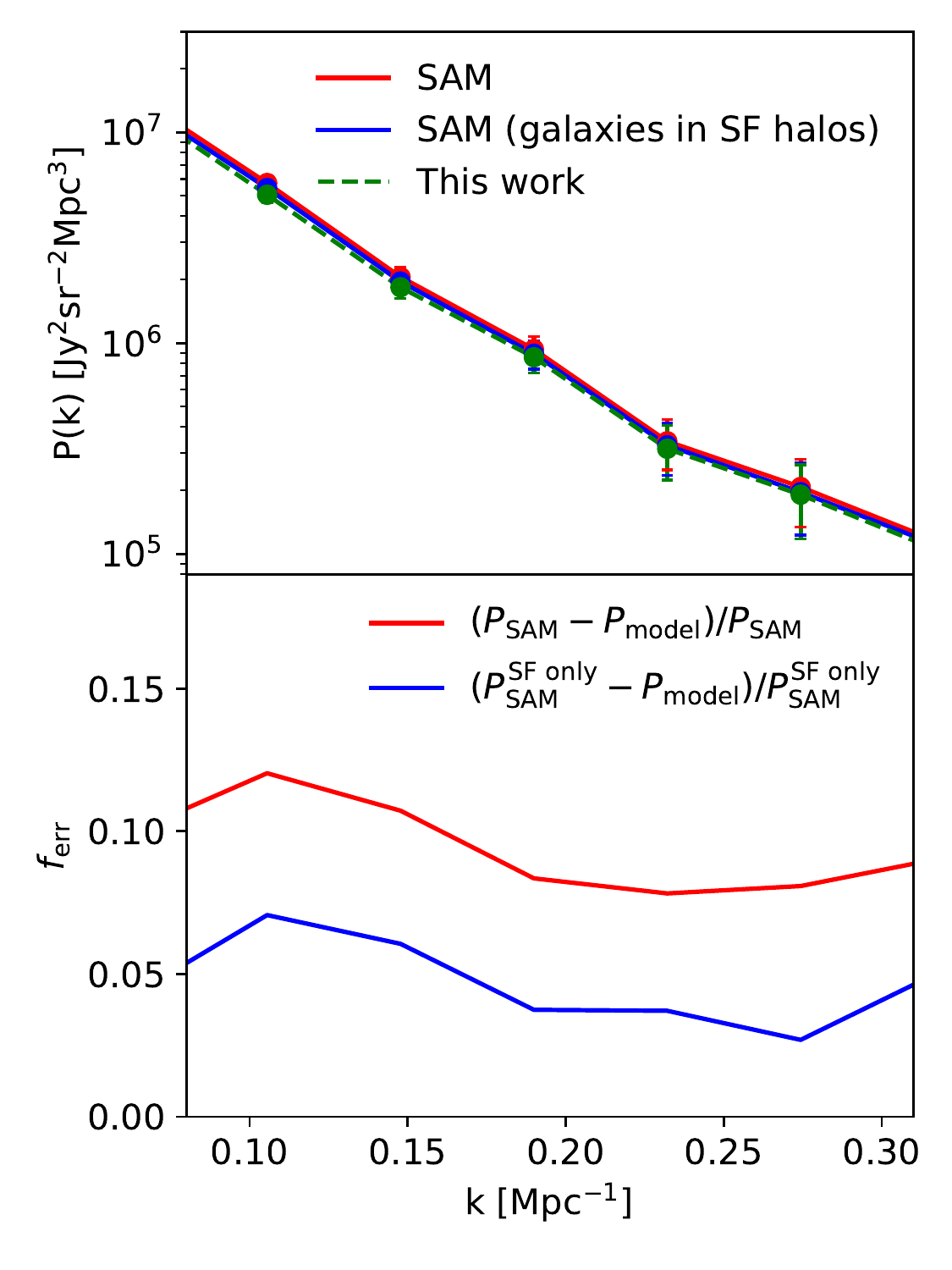}
    \caption{\textit{Top}:Fiducial COMAP CO J=1-0 power spectrum predictions obtained from the sub-mm SAM run within a lightcone extracted from a dissipationless N-body simulation, compared with the prediction from the empirical model presented in this work. The red lines show the predictions from the sub-mm SAM lightcone including all the simulated galaxies, while the blue lines show the lightcone results only including galaxies in star-forming halos. The green dashed line shows the power spectrum predicted by applying the empirical model introduced in this work to the star-forming dark matter halos in the mock lightcone. The error bars show $1-\sigma$ measurement errors typical of the COMAP experiment. \textit{Bottom:} The red curve shows the CO power spectrum fractional difference between sub-mm SAM and this work. The blue curve shows the fractional difference between CO power spectrum predicted by sub-mm SAM where only galaxies living in the star-forming halos are considered, and the power spectrum predicted by this work.}
    \label{fig:COMAP_Pk}
\end{figure}

We present the CO J=1-0 power spectra from the sub-mm SAM lightcone and the empirical model introduced in this work in Figure~\ref{fig:COMAP_Pk}. We show cases in which the sub-mm SAM power spectrum is computed from all galaxies in the lightcone compared with a case where we exclude galaxies in quenched halos (using the same criterion presented in section \ref{sec:model}). We also compare the CO J=1-0 power spectrum using the predictions of sub-mm line luminosities and scatter predicted by the empirical model introduced in section \ref{sec:model}. The error bars show the COMAP $1-\sigma$ measurement error contributed by the instrumental noise, CO J=2-1, and CO J=3-2 interloper lines. We find the amplitude decrease of the power spectrum in the sub-mm SAM caused by ignoring quenched halos is much smaller than the measurement error, indicating that the star forming halos dominate the LIM statistics. This justifies our treatment of quenched halos in the empirical model. The power spectrum predicted by our empirical model has amplitude 8\%-12\% lower than the sub-mm SAM full COMAP power spectrum and 3\%-7\% lower than the SAM star-forming power spectrum. Reduced $\chi^2$ tests show that these three sets of power spectra are indistinguishable considering the measurement error.

\section{Discussion and conclusion}\label{sec:conclusion}
Sub-mm LIM is an emerging observational technique that has the potential to constrain many important features of galaxy evolution and cosmology.   Computationally efficient, physically grounded theoretical sub-mm line emission models that accurately capture the strength, scatter, and correlations of multiple lines are crucial for forecasting for and interpreting the results of these experiments. In this work, we refine the state of the art multi-line sub-mm SAM simulation framework proposed in Popping2019 and construct a simple empirical model calibrated to this physically motivated simulation. Specifically, we model the simulated sub-mm line luminosity versus halo mass relation of star-forming halos as a double power law, and provide carefully calibrated models for the halo duty cycle factor, the dispersion in luminosity at fixed halo mass, and the correlations among different sub-mm lines.

We find that our empirical model can reproduce the integrated line intensity and power spectrum of [\cii], CO J=1-0 to J=5-4, and [\ci] J=1-0 to J=2-1 predicted by the sub-mm SAM simulation with less than 10\% and 25\% fractional error at $z\geq1$. This model provides a very computationally efficient, yet physically grounded approach for CO high J interloper line removal, molecular hydrogen density constraints, and other [\cii] and CO LIM survey forecasts \citep{2021arXiv210614904B,Pullen2022EXCLAIM}. For applications that may require greater accuracy than that of our current empirical model,  we also provide the full tabulated $L(M,z)$, $f_\mathrm{duty}(M,z)$, and the covariances among [\cii], CO, and [\ci] lines from the sub-mm SAM predictions at $0\leq z\leq9$ in \url{ https://users.flatironinstitute.org/~rsomerville/Data\_Release/LIM/}.\par 

A careful comparison between the SAM + sub-mm SAM $L(M,z)$ relations and other models (including \cite{2008A&A...489..489R,2010JCAP...11..016V,2013ApJ...768...15P,2015ApJ...806..209S,2016ApJ...817..169L,2018MNRAS.478.1911P,2019MNRAS.488.3014P}) can be found in \cite{2021ApJ...911..132Y} Section 3.2. As discussed in \cite{2021ApJ...911..132Y}, various $L(M,z)$ models are in reasonable agreement at $11.5\lesssim\log(M/[M_\odot])\lesssim12$, but can be significantly different beyond this halo mass range. This is caused by a lack of sub-mm observational results at very low and high halo masses. The sub-mm line targets selected by galaxy surveys are mainly galaxies with $\mathrm{SFR}\sim10^1-10^2[M_\odot/\mathrm{yr}]$ and halo mass $M\sim10^{11.5}-10^{12}[M_\odot]$. Halos more massive than this range are very rare objects, while galaxies living in halos less massive than this range are generally too faint to be resolved. The upcoming LIM surveys that cover large cosmic volume and faint sub-mm emitter contributions will provide stronger constraints to the shape of $L(M,z)$ relations beyond $11.5\lesssim\log(M/[M_\odot])\lesssim12$. The main advantage of parameterizing $L(M,z)$ relations into double power law is that we can capture the shape of $L(M,z)$ at very low and high halo masses by parameters $\alpha$ and $\beta$, which are sensitive to the strength of stellar feedback and AGN feedback. We leave a study that connects the double power law $L(M,z)$ relation slopes to various feedback strengths to future works.

The empirical models for these emission lines will be tested by upcoming surveys.  For example, preliminary detections of [\cii] and CO intensity \citep{2018MNRAS.478.1911P,2019MNRAS.489L..53Y,Keating2016} using auto- and cross-power spectra have been used along with priors informed by luminosity functions to constrain empirical models for these lines \citep{2018MNRAS.475.1477P,2019MNRAS.488.3014P}.  Upcoming line intensity mapping surveys will be able to measure power spectra and intensities at much higher precision, which will constrain these models even further.  Future works that can connect the parameters in empirical models to physical processes will then be able to measure the physics of galaxies.

We note the major caveats and uncertainties of the models we present here. First, all \emph{a priori} simulation based predictions of sub-mm line emission are sensitive to assumptions about the sub-grid physics incorporated into the simulation, which govern the physics of star formation, stellar feedback, black hole feedback, etc., shaping the star formation history, gas content, and quenching of galaxies. Significant uncertainties in these physical processes and how to implement them through sub-grid prescriptions in cosmological simulations still persist \citep[e.g.][]{SD15}. Moreover, even for a given set of underlying kpc-scale galaxy properties, the sub-mm line predictions are also sensitive to the assumptions about the properties of the ISM (e.g., the mass distribution and radial profiles of molecular clouds), which are in general not directly resolved in these simulations or in observations except for a few very nearby galaxies. The sub-mm SAM approach used here is no exception, but with the advantage that it is computationally efficient enough to explore the sensitivity to these assumptions, as was done extensively in Popping2019. A further limitation of the Popping2019 approach is that \textsc{DESPOTIC} only computes the line emission from gas in molecular clouds. Some of the observed [\cii] emission can also arise from more diffuse cirrus gas, and this is not included in our current models. 

We find the correlation coefficients among CO lines and [\ci] lines simulated by the sub-mm SAM framework agree with the \textsc{EGG} simulation results reported in \cite{2021arXiv210301964S}. The value of $\rho_\mathrm{CO,CI}$ predicted by the sub-mm SAM is slightly lower than \textsc{EGG}'s predictions. This is because \cite{2021arXiv210301964S} only study $\rho_\mathrm{CO,CI}$ up to $z=4$, while this work covers a wider redshift range $0\leq z\leq9$. More specifically, at $z\gtrsim4$, galaxies are more compact and gas rich, and have higher star formation surface density. Since the sub-mm SAM assumes the UV radiation field strength to be proportional to the galaxy star formation rate surface density, carbon atoms in high redshift molecular clouds tend to be mostly ionized, leaving limited material that can produce CO and [\ci] lines. In this case, CO and [\ci] lines can be weakly or even negatively correlated. We also find $\rho_\mathrm{CII,CO}$ and $\rho_\mathrm{CII,CI}$ predicted by the sub-mm SAM are significantly higher than the predictions of the \textsc{EGG} simulator. Considering the fact that the [\cii] fine structure line comes from multiple ISM phases, while CO and [\ci] lines are only emitted by  dense molecular ISM regions, we believe this difference may be caused by different assumptions about the molecular cloud properties implemented in the sub-mm SAM and \textsc{EGG}. The limitations on modeling the ionized gas phase in the sub-mm SAM can also lead to biased estimates of the correlation coefficients between [\cii] and molecular ISM tracers. Empirically constraining these properties would be a valuable target for future LIM cross-correlation observations. However, the sub-mm SAM does successfully reproduce [\cii], CO, and [\ci] observational scaling relations at redshift $0\leq z\lesssim6$, and we emphasize that the simulation results we used for model calibration in this work are consistent with current observations of individual sub-mm sources.\par

An additional limitation is that we do not attempt to model the correlation of the residuals in the $L(M)$ relations with higher order halo properties. For example, for stellar masses and star formation rates, it is known that the residual of these quantities from the mean at a given halo mass is correlated with other halo properties, such as formation history \citep[e.g.][]{Matthee:2017}, which are known to be correlated with the halo clustering strength via a process called \emph{assembly bias} \citep{Croton:2007}. We can think of these as limitations of the standard halo model approach, which could be relatively easily overcome by using SAMs run within merger trees extracted from dissipationless simulations, which we have done in other work (e.g. \cite{2021arXiv211103077G,2021MNRAS.508..698H}). 

Due to the low angular resolution of LIM surveys, interloper lines and continuum foreground contamination, including the Milky Way emission and the cosmic infrared background, are the main challenges for LIM data analysis. Cross-correlating LIM surveys with galaxy surveys that have much higher spatial resolution is one of the best options for removing these contaminants. In this work, we model the aggregate emission of all the galaxies within a halo, and do not attempt to separately model central and satellite galaxies or the spatial distribution of satellite galaxies within halos. However, there is no reason in principle that the SAM approach (which also produces predictions for optical properties of individual galaxies) could not be used to develop joint models for traditional galaxy surveys and LIM surveys. We leave this to a future work. \par

\section{Acknowledgement}
We thank Shenglong Wang for IT support. We thank Dongwoo Chung for useful comments on
a draft manuscript. ARP was supported by NASA under award numbers
80NSSC18K1014 and NNH17ZDA001N. RSS is supported by the Simons Foundation. PCB was supported by the James Arthur Postdoctoral Fellowship.

\bibliography{LM}{} 
\bibliographystyle{aasjournal}

\end{document}